\documentclass[10pt]{article}
\usepackage[utf8]{inputenc}
\usepackage{amsmath, amsthm, amssymb, tikz}
\usepackage{amsfonts}
\usepackage{fullpage}

\usepackage{amsmath,color}

\RequirePackage{youngtab}
\usepackage{amsmath,amssymb,color,graphics,graphicx}
\usepackage{bm}

\def\a{\alpha}
\def\b{\beta}
\def\g{\gamma}
\def\d{\delta}
\def\ep{\varepsilon}

\def\a{\alpha}
\def\b{\beta}

\def\g{\gamma}
\def\d{\delta}
\def\g{\gamma}

  \def\A{{\mathcal{A}}}

\title{Cauchy relations in linear 
elasticity: Algebraic   and  physics aspects
}
\author{Yakov Itin\footnote{Electronic address: itin@math.huji.ac.il, itin@g.jct.ac.il}\\
Jerusalem College of Technology, Jerusalem,  \\ and Institute of Mathematics, The Hebrew University of
  Jerusalem,  Jerusalem, Israel.}
 
\date{\today}

\begin{document}

\maketitle
\begin{abstract}
The Cauchy relations distinguish between rari- and multi-constant linear elasticity theories.  
These relations are treated in this paper in a form that is invariant under two groups of transformations: indices permutation and general linear transformations of the basis. The irreducible decomposition induced by the permutation group is outlined.  The Cauchy relations are then formulated as a requirement of nullification of an invariant subspace.  A successive decomposition under rotation group allows to define the partial Cauchy relations and two types of elastic materials. We explore several applications of the full and partial Cauchy relations in physics of materials. The structure's deviation from the basic physical assumptions of Cauchy's model is defined in an invariant form. The Cauchy and non-Cauchy contributions to Hooke's law and elasticity energy are explained. We identify wave velocities and polarization vectors that are independent of the non-Cauchy part for acoustic wave propagation. Several bounds are derived for the elasticity invariant parameters.

\end{abstract}


\section{Introduction}
In linear elasticity of anisotropic three-dimensional media, the  stress tensor $\sigma^{ij}$ is assumed to be linearly related  to the  strain tensor $\ep_{kl}$. This fact  is  well-known as the anisotropic Hooke's law, see, e.g.,  \cite{Landau}, \cite{Love}, \cite {Marsden}, \cite{Nye}, \cite{Sirotin-book}, 
\begin{equation}\label{Hook}
    \sigma^{ij}=C^{ijkl}\epsilon_{kl}.
\end{equation}
Here, the indices run from 1 to 3. For pairs of repeating indices (one upper and one lower), the Einstein summation rule is assumed. The fourth-order tensor $C^{ijkl}$ is referred to as the {\it elasticity (stiffness)  tensor}. Owing to the definition (\ref{Hook}), the elasticity tensor satisfies the fundamental symmetry relations. In particular, the so-called {\it minor symmetries}, 
\begin{equation}\label{el-sym1}
    C^{ijkl}=C^{jikl}=C^{ijlk}
\end{equation}
follow straightforwardly from the symmetries of the stress and strain tensors,  $\sigma^{ij}=\sigma^{ji}$ and   $\epsilon_{kl}=\epsilon_{lk}$. The {\it major symmetry} of the elasticity tensor,
\begin{equation}\label{el-sym2}
    C^{ijkl}=C^{klij},
\end{equation}
results from an additional condition for the existence of the elasticity energy,
\begin{equation}
    E=\frac 12C^{ijkl}\epsilon_{ij}\epsilon_{kl}\,.
\end{equation}
 As a result, only 21 of the 81 components of the elasticity tensor are linearly independent. According to an algebraic description, this means that in the 81-dimensional vector space of fourth-order tensors, the elasticity tensor forms a subspace of dimension 21. This elasticity tensor subspace is  preserved   under arbitrary invertible transformations of the basis, i.e., it is invariant under the action of the group $GL(3,\mathbb R)$.

In 1827, August Cauchy presented a microscopic model to describe the elasticity phenomena.
The total number of independent components of $C^{ijkl}$ is limited to 15 in this {\it rari-component model} due to additional symmetries (such as the central symmetry of the local force and harmonic potential). 
The tensor $C^{ijkl}$ was supposed to meet six additional independent conditions known as the {\it Cauchy relations} (see \cite{Love}, \cite{Perrin},\cite{Zener1947}):  
\begin{eqnarray}\label{el-Cauchy-1}
C^{1122}&=&C^{1212}\  ;\ \quad C^{1133}=C^{1313}\  ;\ \quad C^{1123}=C^{1213}\nonumber\\
C^{2233}&=&C^{2323}\  ;\ \quad C^{2213}=C^{1223}\  ;\ \quad C^{3312}=C^{1323}.
\end{eqnarray}
In Voigt's notation,  the symmetric pairs of indices are  identified as components of a  6-dimensional vector
\begin{equation}\label{Voigt}
 1\cong 11\,, \,2\cong 22\,,\,3\cong
33\,,\, 4\cong 23\,,\, 5\cong 31\,,\, 6\cong 12\,.
\end{equation}
Then,  the Cauchy relations 
(\ref{el-Cauchy-1})  read
\begin{eqnarray}\label{el-Cauchy-1x}
C^{12}&=&C^{66}\  ;\ \quad C^{13}=C^{55}\  ;\ \quad C^{14}=C^{65} \nonumber\\
C^{23}&=&C^{44}\  ;\ \quad C^{25}=C^{64}\  ;\ \quad C^{36}=C^{54}.
\end{eqnarray}

The debate about the right number of elasticity tensor components raged on until the early twentieth century when the existence of materials with all 21 independent components was empirically justified. As a result, the {\it  multi-component model}, which has 21 independent elasticity tensor components, has been validated.  Besides these experimental findings, theoretical study on the Cauchy relations has taken many different directions:
\begin{itemize}
    \item [(1)] {\it  Analytical derivations: } The precise analytical requirements for confirming the Cauchy relations in the setting of microscopic lattice models have been widely investigated in the mathematical physics literature; see \cite{Love},  \cite{Epstein1946}, \cite{Zener1947}, \cite{MacDonald1972}, \cite{Elcoro2010},\cite{Perrin} and the references given therein. 
    \item [(2)] {\it Algebraic representation:} In the framework of tensor algebra and group theory, the numerous study of the algebraic and symmetry aspects of the elasticity tensor  was provided. These researches was focused on the following set of problems: 
(i) Different types of  decomposition  of the elasticity tensor \cite{Backus},\cite{Baerheim},\cite{Cowin1989},\cite{Cowin1992},\cite{Cowin1995a},\cite{quadr},\cite{Rychlewski1},\cite{Rychlewski2}. 
(ii) Investigation of the precise number of independent  symmetry classes of elasticity tensor, see  \cite{Forte1},\cite{Forte2},\cite{Olive},\cite{Desmorat}. The irreducible portions of the elasticity tensor for different symmetry classes of crystals were reported in  \cite{Itin-MMS}. 
(iii) The algebraic representation of Cauchy relations addresses the issues of their precise covariant form and accurate algebraic interpretation; see \cite{Podio},\cite{Podio1}, \cite{Lancia} \cite{Fosdick},\cite{Cauchy1},\cite{Cauchy2}.   
\item[(3)] {\it Physics conclusions:} The properties of hypothetical materials with elasticity tensors that match precisely the Cauchy relations have been investigated in \cite{Haus},\cite{Cauchy2},\cite{Bona},\cite{Rubin}. 
Because the full Cauchy relations are not exactly fulfilled, it is worthwhile to investigate the deviations of elasticity properties from the ideal Cauchy model for natural materials. In particular, in \cite{Haus} the physics meaning of the deviation term, its magnitude, and its sign are discussed. Another important issue is the application of the Cauchy relations to acoustic wave propagation problems; see \cite{Cauchy2} for some preliminary results.
\end{itemize}

The above-mentioned issues (2) and (3) are addressed in this paper. 
The organization of the paper is as follows: In Section 2, we discuss the geometrical meaning of the Cauchy relations in linear elasticity. We apply the irreducible decomposition of the elasticity tensor under the permutation group $S_4$ and identify the Cauchy and non-Cauchy parts as invariant subspaces. Then the Cauchy relations are expressed as a requirement of nullification of an invariant subspace. Under the rotation group $SO(3, \mathbb R)$, from the Cauchy and non-Cauchy parts,  we obtain successive invariant decompositions into subspaces of dimensions $(1+5+9)$ and $(1+5)$, respectively. Then we define the partial Cauchy relations that mean the nullification of the five dimensional subspace of the non-Cauchy part. With the non-zero scalar invariant at hands we are able to  divide the elastic materials into two disjoint classes.  

In Section 3, we discuss several physics applications of full and partial Cauchy relations. We present the explicit expressions for high-symmetry crystals and provide some examples for two types of materials.  In contrast to the full Cauchy relations,  the partial Cauchy relations  are satisfied for isotropic materials and cubic crystals. We also propose an invariant measure of violation from the ideal Cauchy model. We study the invariant decomposition of Hooke's law. In Section 4, we consider the invariant decomposition of the elasticity energy. The contributions of the Cauchy and non-Cauchy parts are identified and the bounds of the elasticity parameters are derived. In Section 5, we study the implication of the Cauchy and non-Cauchy parts in the propagation of acoustic waves. We define the Cauchy and non-Cauchy parts of the Christoffel tensor and investigate their contributions to acoustic problems. We demonstrate that the velocity of the pure longitudinal wave and the polarization of the pure shear wave are both independent of the elasticity tensor's non-Cauchy part.  In the Conclusion section, we briefly recall the results presented in the paper. 


\section{Cauchy relations---algebraic aspects}
We apply the decomposition of the elasticity tensor into smaller sub-tensors as the first step in understanding the algebraic meaning of the Cauchy relations (\ref{el-Cauchy-1}). In the literature, the harmonic decomposition of the tensor $C^{ijkl}$ under the group of space rotation is commonly used; see, e.g.,  \cite{Backus}. The elasticity tensor is decomposed into five sub-tensors using this method: two scalar tensors, two traceless second-order tensors with five components each, and one traceless fourth-order tensor with nine components. Since the  sub-tensors of  equal dimensions belong to isomorphic subspaces, they can be replaced by arbitrary  independent linear combinations. Consequently, such decomposition is not unique. 
A similar approach can be traced back to the classical Lam\'e's representation of the elasticity tensor of an  isotropic material by two scalars $\lambda$ and $\mu$. For an anisotropic material, an analogous construction was applied in \cite{Cowin1989}. 
When a non-unique decomposition of the elasticity tensor is used,  one cannot expect to achieve a defined physical meaning of specific irreducible parts. 

In this paper, we advocate a {\it two-level decomposition}  strategy as an alternative. Such an approach was initially put forth by Sirotin \cite {Sirotin} {\footnote{We particularly appreciate the anonymous reviewer who provided us with such significant information and accurate references.}} in 1975 in accordance with Weyl's concepts; see \cite{Weyl}, Chapter IV, Section 5. The elasticity tensor is decomposed with regard to the group $GL(3,\mathbb R)$ on the first level.  Then, with respect to the subgroup $O(3,\mathbb R)$, the decomposition of two independent $GL(3,\mathbb R)$-invariant components is constructed. It was underlined by Sirotin that this method differs from the conventional harmonic approach and offers a unique decomposition {\footnote{However, in general, a decomposition of a tensor space under $GL(3,\mathbb R)$ need not be unique. Some explicit examples are given in \cite{IR}}.} In 1988 Mochizuki \cite{Mochizuki} proposes an alternative of the 
harmonic decomposition of the elasticity tensor, which ``is convenient for surface-wave
analysis." Mochizuki arrives at his decomposition also through a two-step procedure. 

Our approach is slightly different.
Since the permutation of indices is involved naturally in the very definition of a tensor, we start with the irreducible decomposition that is invariant under the permutation group $S_4$ instead of the $GL(3,\mathbb R)$ group. In the case of the elasticity tensor, this technique yields an entirely  {\it unique decomposition} of $C^{ijkl}$ into two sub-tensors $S^{ijkl}$ and $A^{ijkl}$ of dimensions 15 and 6, respectively. We show that the Cauchy relations appear just at this level. At the second level, the tensors $S^{ijkl}$ and $A^{ijkl}$ are decomposed into smaller sub-tensors that are invariant under the group of rotations $SO(3, \mathbb R)$. This decomposition is unique as well. The isomorphic subspaces that emerge here cannot be mixed since they belong to two different spaces. Within the two-level approach, we can decompose the Cauchy relations themselves into a system of two sub-relations. Then we are ably to introduce a notion of {\it partial Cauchy relations.}

\subsection{Elasticity tensor and its groups of transformations}
We  start with the basic definition of a tensor as a linear map between vector spaces. Consider a vector space $V$ over the real number field  ${\mathbb R}$. In elasticity theory, the space $V$ is assumed  to be isomorphic to the ordinary 3-dimensional space ${\mathbb R}^3$ (the 3-dimensional translation space of a Euclidean 3-dimensional affine space). A general fourth-order tensor $C$ is defined as a multi-linear map of the Cartesian product of four copies of $V$ into  the  field ${\mathbb R}$, see, e.g., \cite{Marsden}
\begin{equation}\label{Cart-prod}
    C:V\times V\times V\times V\to {\mathbb R}\,.
\end{equation}
Using the basis $e_i$ of the vector space $V$  the tensor $C$ is expressed via the tensor basis $e_i\otimes e_j\cdots$ as 
\begin{equation}\label{Cart-prod1}
    C=C^{ijkl}e_i\otimes e_j\otimes e_k\otimes e_l.
\end{equation}
We observe that the basic definition of a  tensor naturally involves {\it two different groups of transformations:} 
\begin{itemize}
    \item A {\it discrete permutation group}  that is related to the order of the spaces $V$ in the (non-commutative) Cartesian product (\ref{Cart-prod}). For the fourth-order elasticity tensor, it is the permutation (symmetry) group of four distinct elements, denoted by $S_4$. This group is only dependent on the tensor's order. It is unaffected, for example, by the geometric properties of the space $V$, such as metric, dimension, and so on.
    \item A {\it continuous group} of tensor transformations that is generated by the geometric structure defined in the vector space $V$. It is a group of general linear transformations, $GL(3, \mathbb R)$, for a bare vector space without any additional structure like inner product, norm, and so on.
     For a vector space endowed with the metric tensor, it is the group of orthogonal transformations $O(3, \mathbb R)$ with its subgroup of proper rotations $SO(3, \mathbb R)$. This group is the same for all tensors, regardless of their order.
\end{itemize}  
Owing to the  Schur-Weyl duality theorem (see \cite{Weyl}, Chapter III, Section 4),  the decomposition of a  tensor  relative to the symmetry group  is invariant under the action of the general linear group  and its subgroups. Consequently,  the two  groups listed above can be applied  independently. 

\subsection{$S_4$-irreducible parts of the elasticity tensor}
In this section, we describe the $S_4$-invariant sub-tensors of the elasticity tensor. In other words, we are looking for the $S_4$-invariant subspaces of the vector space ${\cal C}$. 
For this problem, the combinatorics decomposition based on  Young’s diagram technique is usually applied  \cite{Hamermesh}.

Due to the basis symmetries (\ref{el-sym1}) and  (\ref{el-sym2}),  the  permutation group decomposition   is described by the following sum of two Young's diagrams: 
\begin{equation}\label{irr3a}
  {\rm SYM}\Big(\Yvcentermath1\yng(2)\, \otimes\, \yng(2)\Big)
  =\yng(4)\,\oplus\, \yng(2,2)\,
\end{equation}
Let us go over the notations used here in a little more detail. The  left-hand side of Eq. (\ref{irr3a}) describes  the general  elasticity tensor   having two symmetric pairs of indices. It is in agreement with the  minor symmetries (\ref{el-sym1}).  This expression is additionally symmetrized with regard to the permutation of two pairs of its indices in order to express the major symmetry (\ref{el-sym2}). This additional symmetrization is denoted by $\rm SYM$. The first diagram on the right-hand side of Eq. (\ref{irr3a}) depicts a fully symmetric tensor, while the second one depicts a  tensor that is partially symmetric and partially skew-symmetric.

According to the diagrams in (\ref{irr3a}), the elasticity tensor is decomposed under the  action of the group $S_4$ into a sum of two  sub-tensors,
 \begin{equation}\label{el-tens-decomp}
C^{ijkl}=S^{ijkl}+A^{ijkl}. 
\end{equation}
   In group theory, the explicit expressions of the corresponding sub-tensors are calculated by applying the products of the symmetrization and antisymmetrization operators from the $S_4$-group algebra. These operations are applied in accordance with the corresponding Young diagrams.  The case of the elasticity tensor is relatively simple, so the two sub-tensors can be obtained straightforwardly. The first  term in (\ref{el-tens-decomp}) is given by the full symmetrization of the tensor in all its four indices. We denote this operation by round parenthesis. Due to the basic symmetries (\ref{el-sym1}) and (\ref{el-sym2}), it is expressed by {\footnote{Equations (11), (12), and (13) first appeared in Backus [2], whose starting point seems to be the harmonic decomposition of (totally) symmetric tensors.}}
 \begin{equation}\label{el-Cauch-part}
S^{ijkl}=C^{(ijkl)}=\frac 13\left(C^{ijkl}+C^{iklj}+ C^{iljk}\right).
\end{equation}
The second irreducible part can be merely considered as a residue term,   $A^{ijkl}=C^{ijkl}-S^{ijkl}$. Thus, it is  given by
\begin{equation}\label{el-non-Cauch-part}
A^{ijkl}=
\frac 13\left(2C^{ijkl}-C^{ilkj}-C^{iklj}\right). 
\end{equation}

The  tensors $S^{ijkl}$ and $A^{ijkl}$ contain 15 and 6 independent components, respectively. These dimensions can be derived by combinatorics  formulae, see, e.g., \cite {IR}, or merely by the direct listing of the independent components, as in  \cite{Cauchy2}. The sub-tensors $S^{ijkl}$ and $A^{ijkl}$ inherit the basic symmetries (\ref{el-sym1}) and (\ref{el-sym2}) of the elasticity tensor, 
\begin{eqnarray}\label{S-symm}
S^{ijkl}=S^{jikl}=S^{ijlk}=S^{klij},
\end{eqnarray}
and
\begin{eqnarray}\label{A-symm}
A^{ijkl}=A^{jikl}=A^{ijlk}=A^{klij}.
\end{eqnarray}
These two tensors also meet the extra symmetry relations listed below:   
\begin{equation}\label{el-non-Cauch-ident}
 S^{(ijkl)}= S^{ijkl}\,,\qquad {\rm and}\qquad  A^{i(jkl)}= 0\,.
\end{equation}

Since the partially symmetric tensor $A^{ijkl}$ has  6 independent components, it can be expressed by a symmetric second-order tensor. We use a definition as in \cite{Itin-MMS}
\begin{equation}\label{def-Delta}
    \Delta_{mn}=\Delta_{nm}=\frac 13 \epsilon_{mil}\epsilon_{njk}A^{ijkl}.
\end{equation}
 Slightly different expressions   were  proposed in  \cite{Haus} and \cite{Cauchy1}. 
In Eq. (\ref{def-Delta}),  $\epsilon_{ijk}$ denotes the permutation Levi-Civita's pseudotensor,
\begin{equation}\label{app1-1}
    \epsilon^{ijk}=\left\{ \begin{array}{rl}
1 & \mbox{if $(ijk)$ is an even permutation of $(123)$};\\
-1 & \mbox{if $(ijk)$ is an odd permutation of $(123)$};\\
0&\mbox{otherwise} 
.\end{array} \right.
\end{equation}
Due to the standard three-dimensional contraction relation,  $\epsilon^{ijk}\epsilon_{imn}=\d^j_m\d^k_n-\d^j_n\d^k_m$, we can invert  equation (\ref{def-Delta}) 
\begin{equation}\label{app1-2}
    A^{ijkl}=\frac 12 \left(\epsilon^{ikm}\epsilon^{jln}+
    \epsilon^{ilm}\epsilon^{jkn}\right)
    \Delta_{mn}. 
\end{equation}
The latter expression is easily verified by straightforward substitution into (\ref{def-Delta}). Explicitly,  the six independent components of  $A^{ijkl}$ are expressed as the components of the symmetric matrix 
\begin{equation}
    A^{ijkl}=\begin{pmatrix}
A^{2233} & A^{3312}& A^{2213}\\
* & A^{1133} & A^{1123}\\
* & * & A^{1122}
\end{pmatrix}=
\left(\begin{array}{rrr}
\Delta_{11} &-\Delta_{12}  & -\Delta_{13}\\
* & \Delta_{22} & -\Delta_{23}\\
* & * & \Delta_{33}
\end{array}	\right).
\end{equation}
In terms of vector space algebra, the decomposition (\ref{el-tens-decomp}) can be explained in more detail.  
Let us denote the vector spaces corresponding to the tensors $C^{ijkl}\,, S^{ijkl}$, and $A^{ijkl}$ as ${\cal C}\,,{\cal S}$, and ${\cal A}$, respectively. The results of the $S_4$-invariant decomposition of the elasticity tensor can be formulated in the algebraic form:
\begin{itemize}
    \item Since the tensors $S^{ijkl}$ and  $A^{ijkl}$ inherit the symmetries of $C^{ijkl}$, they form {\it  subspaces} of the elasticity tensor space, ${\cal S}\in{\cal C} $ and ${\cal A}\in{\cal C} $.
    \item The subspaces ${\cal S}$ and ${\cal A} $, as well as the total vector space ${\cal C}$,  are invariant under the action of the groups $S_4$ and  $Gl(3, \mathbb R)$. 
    \item The subspaces ${\cal S}$ and ${\cal A}$ are {\it disjoint} (they only have the zero vector in common), i.e.,  ${\cal S}\cap {\cal A}=0$. 
    \item The {\it dimensions} of the vector spaces ${\cal C}, {\cal S},$ and ${\cal A}$  are 21, 15, and 6, respectively, i.e., they meet the relation 
    \begin{equation}
        {\rm dim} \,{\cal C}= {\rm dim}\, {\cal S}+ {\rm dim}\, {\cal A}.
    \end{equation}
     \item The space ${\cal C}$ is {\it decomposed uniquely into the direct sum} of two invariant subspaces 
     \begin{equation}
         {\cal C}={\cal S}\oplus {\cal A}.
     \end{equation}
     It means that there are no other non-zero $S_4$-invariant subspaces of ${\cal C}$. 
     \item When the metric structure $g_{ij}$ on the vector space $V$ is involved, two subtensors are mutually orthogonal in the following sense
\begin{equation}
S^{ijkl}A_{ijkl}:=g_{im}g_{jn}g_{kp}g_{lq}S^{ijkl}A^{mnpq}=0\,.
\end{equation}
This fact follows straightforwardly from the permutation properties of the sub-tensors. 
\end{itemize}

In order to highlight the features of the  $SA$-decomposition, we compare it to another widely-used decomposition (see, e.g., \cite{Podio}). In this construction, the elasticity tensor is similarly divided into two independent terms,
\begin{equation}\label{dev1}
    C^{ijkl}=M^{ijkl}+N^{ijkl}\,,
\end{equation}
where
\begin{equation}\label{dev2}
    M^{ijkl}=C^{i(jk)l}\,,\qquad N^{ijkl}=C^{i[jk]l}\,.
\end{equation}
Remember that the symmetric combination of indices is denoted by the round parenthesis, but the skew-symmetric combination is denoted by the square parenthesis. We denote the vector space corresponding to the tensors $M^{ijkl}$ and $N^{ijkl}$ by ${\cal M}$ and ${\cal N}$, respectively. The definitions (\ref{dev1}--\ref{dev2}) yield 
\begin{itemize}
    \item ${\cal M}$ and ${\cal N} $  are vector subspaces of the 81-dimensional space of the general 4-th order tensors. 
    Under the actions of the groups $S_4$ and $Gl(3, \mathbb R)$, these subspaces are invariant. 
    \item The sub-tensors $M^{ijkl}$ and  $N^{ijkl}$ do not inherit the symmetries of the elasticity tensor $C^{ijkl}$ thus they {\it do not form  subspaces} of the elasticity tensor space, ${\cal M}\notin{\cal C} $ and ${\cal N}\notin{\cal C} $. 
    \item These subspaces are {\it disjoint}---they only have the zero vector in common, ${\cal M}\cap {\cal N}=0$. 
    \item The dimensions of the spaces ${\cal M}$ and ${\cal N} $ are 21 and 6, respectively. It means that ${\rm dim}\,{\cal M}={\rm dim}\,{\cal C}$, so  ${\rm dim}\,{\cal C}\ne{\rm dim}\,{\cal M}+{\rm dim}\,{\cal N}$.
    \item Although the sum of the subspaces equals the vector space ${\cal C}={\cal M}+{\cal N}$, the spaces {\it do not provide the direct sum decomposition} ${\cal C}\ne{\cal M}\oplus{\cal N}$.
\end{itemize}
 
 Comparing the properties of the $SA$ and $MN-$decompositions listed above, we conclude that namely $SA$-decomposition must be referred to as the {\it irreducible invariant decomposition} of the elasticity tensor under the actions of the groups $S_4$ and  $Gl(3, \mathbb R)$. Note that this decomposition is unique. 

\subsection{Cauchy relations}
Two sub-tensors $S^{ijkl}$ and $A^{ijkl}$ are completely independent and both inherit the basic minor and major symmetries of $C^{ijkl}$. Thus these sub-tensors can be considered elasticity tensors in their own right. One can even imagine materials in which only one of these sub-tensors is nonzero.

As we will discuss in more detail in the following section, the fully symmetric tensor $S^{ijkl}$ must be regarded as an essential non-zero part of the elasticity tensor.
 Then all we have to do is try to eliminate the mixed-symmetric part $A^{ijkl}$,
\begin{equation}\label{el-Cauchy-3}
  A^{ijkl}= 2C^{ijkl}-C^{ilkj}-C^{iklj}=0.
\end{equation}
 It's worth noting that there are just 6 linearly independent conditions here.  Substituting the values $i,j,\cdots=1,2,3$ we obtain from (\ref{el-Cauchy-3})  the Cauchy relations (\ref{el-Cauchy-1}). Hence Eq. (\ref{el-Cauchy-3})  presents the Cauchy relations in a covariant form. Moreover, this equation  is irreducible under the action of the permutation group $S_4$. It means that when we apply to Eq. (\ref{el-Cauchy-3}) an arbitrary  symmetrization (or skew-symmetrization) operation, the expression on the left-hand side is preserved or vanishes identically. With the symmetric tensor $\Delta_{mn}$ at hand, we get an equivalent compact formulation of the Cauchy relations 
\begin{equation}\label{el-Cauchy-4}
   \Delta_{mn}=0.
\end{equation}

Various compact expressions for the Cauchy relations can be found in the elasticity literature. In \cite{Haus}, for example, they are given as
\begin{equation}\label{Haus-Cauchy}
    C^{iijk} -C^{ijik}=0.\qquad \mbox{--- no summation in $i$.}
\end{equation}
This formula is not invariant under general linear transformations of the basis because the terms on the left-hand side are not tensors.  Another widely-used compact form of the Cauchy relation, see, e.g.,  \cite {Podio}, \cite{Stakgold},
\begin{equation}\label{el-Cauchy-5}
    C^{ijkl}-C^{ikjl}=0
\end{equation}
 is of a proper covariant nature and coincides with the relation $N^{ijkl}=0$ discussed above. 
 This equation does not fulfill the basic symmetries (\ref{el-sym1}) of the elasticity tensor, despite being equivalent to the Cauchy relations (\ref{el-Cauchy-1}). As a result, the relation (\ref{el-Cauchy-5}) does not define a subspace of the elasticity tensor space.

In fact, the criterion of the $S_4$-invariance produces the Cauchy relations nearly exclusively for any arbitrary linear relationship between the components of $C^{ijkl}$.
Let us assume the existence of  a {\it certain  invariant linear relation} between the components of the elasticity tensor. This is in addition to the elasticity tensor's basic minor and major symmetries.
By taking into account the symmetries (\ref{el-sym1}) and (\ref{el-sym2}), we conclude that a most general linear invariant relation between the components of the elasticity tensor has a form
\begin{equation}\label{el-Cauchy-6}
   Z^{ijkl}= \a C^{ijkl}+\b C^{iklj}+\g C^{ilkj}=0\,,
\end{equation}
where $\a,\b,\g$ are  numerical parameters. 
In general, Eq. (\ref{el-Cauchy-6}) is a system of 21 linear equations. 

First, we  apply the full symmetrization operation of $Z^{ijkl}$ to all its four indices. As a result, we obtain a sub-system of 15 linearly independent equations 
\begin{equation}\label{el-Cauchy-7}
    Z^{(ijkl)}= (\a+\b+\g)S^{ijkl}=0\,,
\end{equation}
As we mentioned above,  the fully symmetric part of the elasticity tensor $S^{ijkl}$ is essential and  non-zero. Consequently, Eq. (\ref{el-Cauchy-7}) yields a relation between the parameters $\a=-\b-\g$.  
When this value of $a$ is substituted into Eq. (\ref{el-Cauchy-6}), we get
\begin{equation}\label{el-Cauchy-6a}
    \b\left(C^{iklj}-C^{ijkl}\right)+\g \left(C^{ilkj}-C^{ijkl}\right) =0\,.
\end{equation}
This is the most general form of the Cauchy relations. 
In particular, with the parameters  $ \b=\g=1$ we get back to  Eq. (\ref{el-Cauchy-3}).  Eq. (\ref{el-Cauchy-5})  is also included in the family (\ref{el-Cauchy-6a}) when the parameters $\b=1,\, \g=-1$ are used. Equivalently, Eq. (\ref{el-Cauchy-6a}) can be written in term of the tensor $A^{iklj}$
\begin{equation}\label{el-Cauchy-6b}
    \b\left(A^{iklj}-A^{ijkl}\right)+\g \left(A^{ilkj}-A^{ijkl}\right) =0\,.
\end{equation}
If the parameters $\b,\g$ do not vanish  simultaneously, Eq. (\ref{el-Cauchy-6b}) is equivalent to $\Delta_{mn}=0$. 
This fact can be derived by   multiplying  both sides of Eq. (\ref{el-Cauchy-6b}) by a pair of permutation pseudo-tensors.

Hence we have seen that the Cauchy relation is a result of an  arbitrary non-trivial linear covariant relation $Z^{ijkl}=0$ between the entries of the elasticity tensor. Furthermore, there is an infinite  family  of covariant expressions of the Cauchy relation. 
Since they are all equal to $\Delta_{mn}=0$, they are also all equal to one another.

 However, from an algebraic standpoint, the form (\ref{el-Cauchy-3}) occupies a distinct position in this infinite family. Indeed, the equations from the set (\ref{el-Cauchy-6b}) define various subspaces of dimension 6 in the 81-dimensional space of fourth-order tensors for different values of the parameters $\b,\g$. Even still, there is only one of these subspaces that belongs to the elasticity tensor's 21-dimensional space.
  This subspace is defined by Eq. (\ref{el-Cauchy-3}).
With this fact in hand, we can formulate the Cauchy  relations (\ref{el-Cauchy-1}) in an invariant algebraic form:

\vspace{0.2cm}

\vspace{0.15cm}

    \fbox{\begin{minipage}
{\dimexpr\textwidth-25\fboxsep-25\fboxrule\relax}
{\bf Cauchy relations: }
The invariant subspace  ${\cal A}$  of  the  elasticity   space ${\cal C}$ is  nullified.

\vspace{0.15cm}
\end{minipage}}

\subsection{$SO(3,\mathbb R)$-decomposition of the elasticity tensor}
According to the irreducible decomposition method, Cauchy relations are characterized by the disappearance of the $S_4$-invariant subspace ${\cal{A}}\subset  {\cal{S}}$. Recall that the vector space ${\cal{C}}$ has only two non-trivial subspaces, ${\cal{S}}$  and ${\cal{A}}$, which are invariant under the group $GL(3,\mathbb R)$. Under a smaller subgroup $SO(3,\mathbb R)$ of  spatial  rotations, an additional decomposition of the elasticity tensor  space is available. Due to Weyl \cite{Weyl}, Chapter V, such decomposition is directly connected  to the metric tensor $g_{ij}$. Formally, it can be treated as a decomposition of a pair of tensors $\{C^{ijkl}\,, g_{ij}\}$ rather than of the elasticity tensor $C^{ijkl}$ itself. Applying the Euclidean metric tensor $g_{ij}={\rm diag}(1,1,1)$, the trace of a tensor can be expressed in a covariant form. Then we are able to identify the trace and traceless  subspases. 

\subsubsection{Cauchy part} Let us start with the fully symmetric part $S^{ijkl}$. By contracting this quantity with the metric tensor, we can define a unique scalar and a unique traceless second-order tensor. Define a scalar $S$ and a tensor  $P^{ij}$ as follows:  
\begin{equation}
    S=g_{ij}g_{kl}S^{ijkl}\,,
\end{equation}
and  
\begin{equation}
    P^{ij}=g_{kl}S^{ijkl}-\frac 13 Sg^{ij}\,.
\end{equation}
Observe that the latter tensor is traceless $P^{ij}g_{ij}=0$. 
Then two irreducible parts of the tensor $S^{ijkl}$  are expressed as 
\begin{equation}\label{S1}
    {}^{(1)}\!S^{ijkl}=\frac 15 Sg^{(ij}g^{kl)}=\frac S{15}\left(g^{ij}g^{kl} +g^{ik}g^{jl}+g^{il}g^{jk}\right).
  \end{equation}
 and
\begin{equation} \label{S2}
 {}^{(2)}\!S^{ijkl}=\frac 67 P^{(ij}g^{kl)}=\frac 17\left(P^{ij}g^{kl}+P^{ik}g^{jl}+P^{il}g^{jk}+P^{jk}g^{il}+P^{jl}g^{ik}+P^{kl}g^{ij}\right).
\end{equation}
Denote the residue part as
\begin{equation}
    R^{ijkl}=S^{ijkl}-{}^{(1)}\!S^{ijkl}-{}^{(2)}\!S^{ijkl}.
\end{equation}
The coefficients chosen in (\ref{S1}) and (\ref{S2}) ensure that the tensor $R^{ijkl}$ is fully symmetric and traceless, as shown in \cite{quadr}. Hence, it is the third irreducible component of $S^{ijkl}$ 
\begin{equation}
    {}^{(3)}\!S^{ijkl}=R^{ijkl}\,.
\end{equation}
As a result, the Cauchy part of the elasticity tensor is decomposed into three irreducible pieces, each of which is invariant under the rotation group $SO(3,\mathbb R)$,  
\begin{equation}\label{S-decomp}
S^{ijkl}={}^{(1)}\!S^{ijkl}+{}^{(2)}\!S^{ijkl}+{}^{(3)}\!S^{ijkl}\,.
\end{equation}
In terms of vector spaces, this decomposition is presented as  a direct sum of three subspaces of the space ${\cal S}$
\begin{equation}\label{S-decomp-S}
{\cal S}={}^{(1)}\!{\cal S}\oplus{}^{(2)}\!{\cal S}\oplus{}^{(3)}\!{\cal S}\,.
\end{equation}
The dimensions of the space ${\cal S}$ is distributed between its subspaces, respectively, as 
\begin{equation}
    15=1+5+9.
\end{equation}
The sub-tensors in (\ref{S-decomp}) are mutually orthogonal. It means that for $A,B=1,2,3$ with $A\ne B$, we have
\begin{equation}
    {}^{(A)}\!S_{ijkl}\quad\!\!\!\! {}^{(B)}\!S^{ijkl}=0.
\end{equation}
These relations follow from the fact that   the tensors $P^{ij}$ and $R^{ijkl}$ are traceless and symmetric. 
\subsubsection{Non-Cauchy part}
For dealing with the non-Cauchy part  $A^{ijkl}$, we define a unique linearly independent scalar invariant 
\begin{equation}\label{A-scalar}
    A=g_{ij}g_{kl}A^{ijkl}.
\end{equation}
In terms of Voigt's notation it reads
\begin{equation}\label{A-scalar0}
    A=\frac 43\left[\left(C^{12}-C^{44}\right)+\left(C^{13}-C^{55}\right)+\left(C^{23}-C^{66}\right)\right].
\end{equation}
Then we construct a fourth-order tensor 
\begin{equation}\label{A1-tens}
{}^{(1)}\!A^{ijkl}=
\frac{A}{12} \left(2g^{ij}g^{kl}-g^{il}g^{jk}-g^{ik}g^{jl}\right) \,,
\end{equation}
that satisfies the symmetry identities of the non-Cauchy part (\ref{A-symm}) and (\ref{el-non-Cauch-ident}). 
The leading coefficient in Eq. (\ref{A1-tens}) is chosen in such a way that the equation $A=g_{ij}g_{kl}{}^{(1)}\!A^{ijkl}$ holds. 
Now we are able to define the decomposition of the non-Cauchy tensor into the scalar and traceless sub-tensors  
\begin{equation}\label{A-decomp}
    A^{ijkl}={}^{(1)}\!A^{ijkl}+{}^{(2)}\!A^{ijkl}\,.
\end{equation}
 The corresponding subspaces intersect only at zero. The symmetries of the sub-tensors ${}^{(1)}A^{ijkl}$ and  ${}^{(2)}A^{ijkl}$ are the same as those of the tensor $A^{ijkl}$. As a result, they constitute invariant subspaces of the $\cal A$ space:    Consequently we are dealing with the direct sum decomposition
\begin{equation}
    {\cal A}={}^{(1)}\!{\cal A}\oplus{}^{(1)}\!{\cal A},\qquad 6=1+5.
\end{equation}
To demonstrate that this decomposition is irreducible, 
the matrix representation $\Delta_{ij}$ is useful.  We can define a unique scalar trace $\Delta=g^{ij}\Delta_{ij}$ 
and then split  the matrix $\Delta_{ij}$  into  the trace  and the traceless pieces, 
\begin{equation}\label{Delta-decomp}
    \Delta_{ij}=\frac 13g_{ij}\Delta +Q_{ij}\,,
    \end{equation}
    where the residue part $Q_{ij}$ satisfies the traceless  relation $Q_{ij}g^{ij}=0$. 
 In terms of the second-order tenors the sub-tensors of the non-Cauchy part are expressed as (\ref{A1-tens}) with $A=2\Delta$.

The second sub-tensor reads 
 \begin{equation}\label{A2-ep}
{}^{(2)}\!A^{ijkl}=\frac 12 \left(\epsilon^{ikm}\epsilon^{jln}+\epsilon^{ilm}\epsilon^{jkn}\right)Q_{mn} \,,
\end{equation}
Its explicit expression is given by 
\begin{equation}\label{A2-g}
{}^{(2)}\!A^{ijkl}=\frac 12 \left(
g^{il}Q^{jk}+g^{ik}Q^{jl}+g^{jl}Q^{ik}+g^{jk}Q^{il}-2g^{ij}Q^{kl}-2g^{kl}Q^{ij}\right)\,,
\end{equation}

The decomposition (\ref{Delta-decomp}) of the symmetric second-order tensor  $\Delta_{ij}$ is  irreducible and  unique. Since this tensor is completely equivalent to the fourt-order tensor $A^{ijkl}$, the decomposition (\ref{A-decomp}) is irreducible and  unique also.

\subsubsection{Two-level decomposition} We described the two-level decomposition of the elasticity tensor. At the first level, the tensor $C^{ijkl}$ is decomposed uniquely into a sum of two subtensors---the Cauchy part $S^{ijkl}$ and the non-Cauchy part $A^{ijkl}$. These sub-tensors are  invariant simultaneously  under the $S_4$-permutations of the indices and  under the  general linear transformation of the basis, $GL(3,{\mathbb R})$. At the second level, the tensors $S^{ijkl}$ and $A^{ijkl}$ are decomposed uniquely under the group of spatial rotations, $SO(3,{\mathbb R})$. As a result, two levels of the decomposition are presented as
\begin{eqnarray}\label{decomp-C-fin}
C^{ijkl}&=&S^{ijkl}+A^{ijkl}\nonumber\\
&=&\left({}^{(1)}\!S^{ijkl}+{}^{(2)}\!S^{ijkl}+{}^{(3)}\!S^{ijkl}\right)+\left({}^{(1)}\!A^{ijkl}+{}^{(2)}\!A^{ijkl}\right).
\end{eqnarray}
In terms of the corresponding vector spaces, we have a unique two-level decomposition 
\begin{eqnarray}
{\cal C}&=&{\cal S}\oplus{\cal A} \nonumber\\
&=&\left({}^{(1)}\!{\cal S}\oplus{}^{(2)}\!{\cal S}\oplus{}^{(3)}\!{\cal S}\right) \oplus\left({}^{(1)}\!{\cal A} \oplus{}^{(2)}\!{\cal A}\right)
\end{eqnarray}
with the dimensions  distributed, respectively, as 
\begin{equation}
    21=15+9=(1+5+9)+(1+5)\,.
\end{equation}
\subsection{Uniqueness of the elasticity tensor decomposition} 
The decomposition (\ref{decomp-C-fin}) is constructed by two scalars, $S$ and $A$, two second-order traceless tensors, $P_{ij}$ and $Q_{ij}$, and a  fully symmetric and  traceless   fourth-order tensor $R_{ijkl}$. 
The same kinds of tensors  arise in the frequently used {\it harmonic decomposition of the elasticity tensor}; see, e.g., \cite{Backus}, \cite{Baerheim}, \cite{Cowin1989},  \cite{Cowin1992}, \cite{Cowin1995a}, \cite{Forte1}, and  \cite{Forte2}.
The latter  decomposition is generated from  the harmonic polynomials, i.e.,  the polynomial solutions of the Laplace equation. The corresponding tensors are required to be completely symmetric and totally traceless. 
The most concise expression of this type was proposed by Cowin \cite{Cowin1989},  
\begin{eqnarray}\label{id1}
C^{ijkl}&=&ag^{ij}g^{kl}+b\left(g^{ik}g^{jl}+g^{il}g^{jk}\right)
+\left(g^{ij}{\hat A}^{kl}+g^{kl}{\hat A}^{ij}\right) +\nonumber\\
&&\left(g^{ik}{\hat B}^{jl}+g^{il}{\hat B}^{jk}+g^{jk}{\hat B}^{il}+g^{ik}{\hat B}^{jl}\right)+Z^{ijkl}\,.
\end{eqnarray}
The first two terms here repeat the well-known Lam\'e's description of the elasticity tensor for isotropic materials. Observe that the five individual terms in in (\ref{id1}) are reducible because they do not meet the basic symmetries of the elasticity tensor. 

An alternative presentation due to Backus   \cite{Backus},  
\begin{eqnarray}\label{id2}
C^{ijkl}&=&H^{ijkl}+ \left(H^{ij}g^{kl}+H^{ik}g^{jl}+H^{il}g^{jk}+H^{jk}g^{il}
+H^{jl}g^{ik}+H^{kl}g^{ij}\right)+\nonumber\\
&&H\left(g^{ij}g^{kl}+g^{ik}g^{jl}+g^{il}g^{jk}\right)+h\left(g^{ij}g^{kl}-\frac12g^{il}g^{jk}-\frac12 g^{ik}g^{jl}\right)+\nonumber\\
&&
\left(h^{ij}g^{kl}+h^{kl}g^{ij}-\frac 12 h^{jl}g^{ik}-
\frac 12h^{ik}g^{jl}-\frac 12h^{jk}g^{il} -\frac 12h^{il}g^{jk} \right),
\end{eqnarray}
does not suffer from these restrictions. In fact, it is equivalent (up to the leading coefficients) to the result of our two-level  decomposition (\ref{decomp-C-fin}). 
However, it is not unique when the decomposition (\ref{id2}) is solely studied on a  $SO(3,\mathbb R)$-invariant level. This is due to the isomorphism of the tensor spaces of the same dimension. In particular, the one-dimension spaces in (\ref{id2}) presented by the scalars $h$ and $H$ are isomorphic to one another. Thus they can be mixed into two arbitrary independent linear combinations as in (\ref{id1}). Similarly, the five-dimensional spaces, presented by the tensors $h^{ij}$ and $H^{ij}$ are isomorphic to one another. Thus they can be rearranged into two other spaces presented in (\ref{id1}) by the tensors ${\hat A}^{ij}$ and ${\hat B}^{ij}$.

The uniqueness of the sub-tensors is a highly sought quality since only unequally defined  tensors may be given an {\it invariant physical sense}. 
The two-level decomposition overcomes the problem of uniqueness. In particular, the decomposition (\ref{decomp-C-fin}) presents the one-dimensional subspaces ${}^{(1)}\!{\cal S}$ and ${}^{(1)}\!{\cal A}$ as originating from separate sources, namely the Cauchy and non-Cauchy spaces ${\cal S}$ and ${\cal A}$, respectively. As a result, even though they have the same algebraic dimensions, these subspaces cannot be combined into a new pair of 1-dimensional subspaces. Similarly, the five-dimensional subspaces ${}^{(2)}\!{\cal S}$ and ${}^{(2)}\!{\cal A}$ cannot be mixed into independent linear combinations, even if they are isomorphic.  In this sense, the sub-tensors given in (\ref{decomp-C-fin}) are defined uniquely.

Consider a simple electromagnetic analogy: Electric and magnetic fields belong to three-dimensional vector spaces. These spaces are algebraically isomorphic one to another. The fields, on the other hand, cannot be mixed, i.e. rearranged into two independent linear combinations. The reason for this is that the electric and magnetic vector fields have distinct origins---twisted and untwisted tensors, respectively. See  \cite{Hehl-Obukhov(2003)} for more details.

\subsection{Partial Cauchy relations---two classes of materials}
Under the action of the rotation group, the 6-dimensional vector space  ${\cal A}$ is decomposed  into the direct sum of two proper subspaces: the 1-dimensional subspace of scalar (trace) tensors ${}^{(1)}\!{\cal A}$ and the   5-dimensional subspace of traceless tensors ${}^{(2)}\!{\cal A}$. 
Thus the Cauchy relations $A^{ijkl}=0$ also can  be divided into a system of two independent equations: 
\begin{equation}\label{Cauchy-full}
    {}^{(1)}\!A^{ijkl}=0\,,\quad{\rm and}\quad {}^{(2)}\!A^{ijkl}=0\,.
\end{equation}
We may now suggest a way to relax the  classical Cauchy conditions by introducing  a notion of {\it partial Cauchy relations}. In particular, we can consider an invariant relations  that nullifies only one subspace of the tensor $A^{ijkl}$. 
We define the {\it partial Cauchy relations} as 
\begin{equation}\label{Cauchy-part1}
    {}^{(1)}\!A^{ijkl}\ne 0\,,\quad{\rm but}\quad {}^{(2)}\!A^{ijkl}= 0\,.
\end{equation}
This is a system of 5 independent equations. This  partial Cauchy relation can be written in terms of the traceless second-order tensor 
\begin{equation}
    Q_{ij}=0 \,.
\end{equation}
We have the following expression (\ref{A-scalar0}) for the scalar $A$. 
\begin{equation}\label{A-scalar1}
A=(4/3)\left(C^{12}+C^{13}+C^{23}-C^{44}-C^{55}-C^{66}\right).
\end{equation}
Then the  
components of the tensor $Q^{ij}$ are expressed as  
\begin{eqnarray}
Q^{11}&=&(4/3) \left(C^{23}-C^{44}\right)-(1/3) A,\qquad Q^{12}= C^{45}-C^{36}, \\
Q^{22}&=&(4/3) \left(C^{13}-C^{55}\right)-(1/3) A,\qquad  Q^{13}=  C^{46}-C^{25},\\
Q^{33}&=&(4/3) \left(C^{12}-C^{66}\right)-(1/3) A,\qquad Q^{23}= C^{56}-C^{14}.
\end{eqnarray}\label{PartCauchy}
Thus the partial Cauchy relations are given by the system 
\begin{eqnarray}
C^{23}-C^{44}=(1/4) A,\qquad C^{45}-C^{36}=0, \\
C^{13}-C^{55}=(1/4) A,\qquad    C^{46}-C^{25}=0,\\
C^{12}-C^{66}=(1/4) A,\qquad  C^{56}-C^{14}=0.
\end{eqnarray}
Due to the relation $Q^{11}+Q^{22}+Q^{33}=0$, we have here only five independent equations. 

The analysis of the irreducible parts of crystals \cite{Itin-MMS} shows that the isotropic media and the cubic class crystals satisfy the partial Cauchy relation of this type.  
We get to the conclusion that the scalar $A$ is crucial, and the partial Cauchy relation (\ref{Cauchy-part1}) is worth to consider. The scalar  relation $A\ne 0$ is the only one that holds for any natural material. 
Using the non-zero scalar $A$, we are now in a position to divide the variety of elastic materials into two distinct classes: 
\begin{itemize}
    \item  The {\it  $A^+$-materials} with a positive scalar parameter $A$,
     \item  The {\it $A^-$-materials} with a negative scalar parameter $A.$
\end{itemize}
We demonstrate in the following section that these two categories of materials do indeed have certain unique physical characteristics.

From the pure algebraic view, we can try to require nullification of the scalar sub-space ${}^{(1)}{\cal A}$ only, i.e. to define the {\it partial Cauchy relations of the second type}: 
\begin{equation}\label{Cauchy-part2}
    {}^{(1)}\!A^{ijkl}= 0\,,\quad{\rm but}\quad {}^{(2)}\!A^{ijkl}\ne 0\,.
\end{equation}
We have here only one condition on the components of the elasticity tensor.
To our knowledge, there are no materials that satisfy the partial Cauchy relation of this type.

\subsection{Results}
Let us briefly recall the results of the algebraic consideration. 
A two-level decomposition of the elasticity tensor was considered. Two distinct pieces—Cauchy parts and non-Cauchy parts—are identified at the first level. These parts are split into rotational covariant sub-tensors at the second level. It's worth noting that the Cauchy and non-Cauchy parts are structurally similar. Each of them contains both the 1-dimensional subspaces 
\begin{equation}
    {}^{(1)}\!S^{ijkl}\sim S\,,\qquad {}^{(1)}\!A^{ijkl}\sim A\,,
\end{equation}
and the 5-dimensional subspaces 
\begin{equation}
    {}^{(2)}\!S^{ijkl}\sim P_{ij}\,,\qquad {}^{(2)}\!A^{ijkl}\sim Q_{ij}\,.
\end{equation}
The Cauchy part contains an additional  9-dimensional subspace ${}^{(3)}\!S^{ijkl}\sim R_{ijkl}$. 
According to the analysis of the crystal systems \cite{Itin-MMS}, the crystal symmetry cannot distinguish between the Cauchy and non-Cauchy parts. The matrices $P_{ij}$ and $Q_{ij}$, in particular, have the same symmetry  structure. They are even proportional to one another in high symmetry systems. In all cases, the {\it algebraic differences } between two components of the elasticity tensor must result in some differences in their physical properties. In the next sections, we  will discuss how these {\it physical differences} can be recognized.
\section{Cauchy relations---physics  aspects}
Owing to the algebraic invariance of the  Cauchy relations, we conclude that they  can have a defined physical meaning even without being strongly fulfilled. In this section, we examine certain physical issues where Cauchy relations  show up. 

\subsection{Cauchy relations for natural materials}
We start with a brief account of the Cauchy and non-Cauchy parts of the isotropic materials as well as of some high-symmetric crystals.
\subsubsection{Isotropic materials}
Since the spatial directions have no influence on the elasticity properties of the isotropic material, only the scalar invariants of $C^{ijkl}$ and the metric tensor $g_{ij}$ may be used to define the elasticity tensor.
 Usually, the tensor $C^{ijkl}$ of the isotropic material is described  by  two Lam\'e moduli $\lambda$ and $\mu$, see e.g. \cite{Landau}, \cite{Love}, as 
\begin{equation}\label{iso}
C^{ijkl}= \lambda\,g^{ij}g^{kl}+\mu\left(g^{ik}g^{lj}+g^{il}g^{jk}
  \right).
\end{equation}
Observe that the two terms here are reducible under the action of the permutation group. 
  The irreducible decomposition of $C^{ijkl}$ is provided by two other tensors, which are independent linear combinations of the terms  in (\ref{iso})
\begin{equation}\label{first''}
   S^{ijkl}= \frac S{15}\left(g^{ij}g^{kl}+ g^{ik}g^{lj}+g^{il}g^{jk}\right), \quad {\rm where}\quad S=5(\lambda+2\mu),
\end{equation}
and
\begin{equation}\label{second'}
 A^{ijkl}=  \frac{A}{12}
\left(2g^{ij}g^{kl}-g^{ik}g^{lj}-
g^{il}g^{jk}\right), \quad {\rm where}\quad A=4 (\lambda-\mu). 
\end{equation}
As a result, the isotropic material's elasticity tensor is expressed as the sum of its two irreducible scalar parts—--one Cauchy and one non-Cauchy. These parts are also rotational invariant, thus we have a decomposition 
\begin{equation}
   C^{ijkl}={}^{(1)}\!S^{ijkl}+{}^{(1)}\!A^{ijkl}.
\end{equation}
The Cauchy relations are equivalent to  one scalar equation 
\begin{equation}
   {}^{(1)}\!A^{ijkl}=0\qquad \Longrightarrow\qquad C^{12}-C^{44}=0.
\end{equation}
In term of the Lam\'e moduli, it reads $\lambda=\mu$. In terms of the Poisson ratio, it yields the classical critical value
\begin{equation}
    \nu=\frac {\lambda}{2\lambda+2\mu}=0.25\,.
\end{equation}

Note that the partial Cauchy relations, $A\ne 0$ and  $Q_{ij}=0$, hold identically for all isotropic materials. Then the non-zero parameter $A$ allows for a covariant  identification of {\it two types of isotropic materials:} 
\begin{itemize}
    \item The  $A^+$-materials with $\lambda>\mu$, or, equivalently,  $\nu>0.25$;\\
    Examples: Gold---$\nu=0.43$, Aluminum---$\nu=0.32$, Copper	---$\nu=0.355$.
    \item The $A^-$-materials with $\lambda<\mu$, or, equivalently,  $\nu<0.25$;\\ 
    Examples: Concrete---$\nu=0.2$, Cast Iron---$\nu=0.24$, Glass---$\nu=0.22$.
\end{itemize}
It is worthwhile to note the recently discussed Poisson's ratio lower bound for isotropic materials, $\nu>0.2$ (see \cite{Mott}).

\subsubsection{Cubic crystals}
With three independent elasticity modules $C^{11}, C^{12}$, and $C^{44}$, the cubic system represents the simplest anisotropic material.  In Voigt's notation, the elasticity modules are arranged into a $6\times 6$ matrix as  follows:
\begin{equation}\label{cub0}
C^{ijkl}= \begin{bmatrix}
  C^{11} & C^{12} & C^{12} & 0 & 0 & 0 \\
* & C^{11} & C^{12} &0 &0 & 0 \\
* & * & C^{11} & 0 & 0 & 0 \\
* & * & * & C^{44} & 0 & 0 \\
* & * & * & * & C^{44} & 0 \\
* & * & * & * & * & C^{44}\, \end{bmatrix}.
\end{equation}
The Cauchy part of the elasticity tensor is presented by two sub-tensors, see \cite{Itin-MMS},  
\begin{equation}
    S^{ijkl}={}^{(1)}S^{ijkl}+{}^{(3)}S^{ijkl}. 
\end{equation}
There is only one independent component in each of these sub-tensors. In (\ref{S1}), however, only the isotropic sub-tensor ${}^{(1)}S^{ijkl}$  is represented as a real scalar. In the case of the cubic crystal, the second part ${}^{(3)}S^{ijkl}$ can be expressed as a dimensionless tensor multiplied by the scalar $R=\sqrt{R_{ijkl}R^{ijkl}}$. 

The non-Cauchy part of the cubic class elasticity tensor contains the first scalar part only
\begin{equation}
    A^{ijkl}={}^{(1)}\!A^{ijkl}.
\end{equation}
This tensor depends of only one scalar invariant $A$. 
Consequently, the elasticity tensor is represented as 
\begin{eqnarray}\label{decomp}
C^{ijkl}&=&S^{ijkl}+A^{ijkl}\nonumber\\
&=&\left({}^{(1)}S^{ijkl}+{}^{(3)}S^{ijkl}\right)+\left({}^{(1)}\!A^{ijkl}\right).
\end{eqnarray}
Three corresponding tensor spaces are one-dimensional.
Cauchy relations are equivalent to only one scalar equation, just like isotropic material 
\begin{equation}
   {}^{(1)}\!A^{ijkl}=0\qquad \Longleftrightarrow\qquad C^{12}-C^{44}=0.
\end{equation}
The partial Cauchy relation is fulfilled for all cubic system materials. 
We identify {\it two types of cubic materials.}

$\bullet$ The $A^+$ cubic materials  satisfy the relation $C^{12}>C^{44}$. 

$\bullet$ The $A^-$ cubic materials  satisfy the relation $C^{12}<C^{44}$.

\noindent  Some examples from \cite{Kaxiras} of $A^+$ and $A^-$ cubic materials  are  given in Tables 1, 2.


\begin{table}[ht]
\begin{minipage}{.45\textwidth} %

    \centering
    \begin{tabular}{l|c|c|c}
    Crystal& $C^{11}$& $C^{12}$ &$C^{44}$\\
    \hline
   AlSb & 0.894& 0.443& 0.416  \\
    InP & 1.022& 0.576& 0.460\\
    InAs & 0.83& 0.453& 0.396\\
    W & 5.224& 2.044& 1.608\\
    Mo & 4.637& 1.578& 1.092
    \end{tabular}
    \caption{The elasticity constants for  $A^+$ \\ cubic materials}

\end{minipage} %
\begin{minipage}{.45\textwidth} %
  \centering
    \begin{tabular}{l|c|c|c}
    Crystal& $C^{11}$& $C^{12}$ &$C^{44}$\\
    \hline
    C & 10.76 &1.250& 5.760\\
    Si& 1.658& 0.639& 0.796\\
    Ge & 1.284& 0.482& 0.667\\
    Ir & 5.800& 2.420& 2.560\\
    Cr & 3.398& 0.586& 0.990
    \end{tabular}
   \caption{The elasticity constants for  $A^-$ cubic materials}
\end{minipage}
\end{table}

\subsubsection{Transversely isotropic system}
The transversely isotropic (hexagonal)   system is characterized by five independent elasticity parameters. 
In Voigt's notation, the elasticity tensor is presented \cite{Landau} by a symmetric  $6\times 6$ matrix,
\begin{equation}\label{hex1}
C^{ijkl}= \begin{bmatrix}
  C^{11} & C^{12} & C^{13} & 0 & 0 & 0 \\
* & C^{11} & C^{13} &0 &0 & 0 \\
* & * & C^{33} & 0 & 0 & 0 \\
* & * & * & C^{44} & 0 & 0 \\
* & * & * & * & C^{44} & 0 \\
* & * & * & * & * & C^{66}\, \end{bmatrix}
\end{equation}
with an additional linear relation 
\begin{equation}
    C^{66}=\frac 12 (C^{11}-C^{12}).
\end{equation}
The two-level decomposition of this elasticity tensor is presented by
\begin{eqnarray}\label{decomp-1}
C^{ijkl}&=&S^{ijkl}+A^{ijkl}\nonumber\\
&=&\left({}^{(1)}S^{ijkl}+{}^{(2)}S^{ijkl}+{}^{(3)}S^{ijkl}\right)+\left({}^{(1)}A^{ijkl}+{}^{(2)}A^{ijkl}\right).
\end{eqnarray}
All the five tensor spaces in the last line are one-dimensional. The two isotropic sub-tensors ${}^{(1)}S^{ijkl}$ and ${}^{(1)}A^{ijkl}$, in particular, are represented by two linear invariants $S$ and $A$, respectively. The three remaining anisotropic parts can be written as scalars multiplied by dimensionless tensors using quadratic invariants, see \cite{quadr}. In particular, 
\begin{equation}
    {}^{(2)}\!S^{ijkl}\sim P=\sqrt{P_{ij}P^{ij}},\qquad {}^{(2)}\!A^{ijkl}\sim Q=\sqrt{Q_{ij}Q^{ij}},\qquad
    {}^{(3)}\!S^{ijkl}\sim R=\sqrt{R_{ijkl}R^{ijkl}}.
\end{equation}

The classical Cauchy relations are expressed by two  equations 
\begin{equation}
    {}^{(1)}\!A^{ijkl}=0\qquad{\rm and}\qquad {}^{(2)}\!A^{ijkl}=0.
\end{equation}
In terms of elasticity invariants, we only have two independent scalar conditions on the elements of the elasticity tensor, 
\begin{equation}
    A=0\qquad{\rm and}\qquad Q=0.
\end{equation}
The partial Cauchy relations discussed above require a weaker condition  $Q=0$. In this case, the invariant $A$ is not zero so we can define two types of transversely isotropic materials:

$\bullet$ The $A^+$ transversely isotropic materials  satisfy the relation 
\begin{equation}
    C^{11}+2C^{13}-2C^{44}-3C^{66}>0
\end{equation}

$\bullet$ The $A^-$ transversely isotropic materials  satisfy the relation 
\begin{equation}
    C^{11}+2C^{13}-2C^{44}-3C^{66}<0
\end{equation}
\subsection{Deviation from the Cauchy model}
Even though no naturally occurring material perfectly satisfies the conventional Cauchy relations, it can be fascinating to look into how much a material's elasticity tensor deviates from the ideal Cauchy model.
 A detailed discussion of this problem is given in  \cite{Haus} and the references therein.  Although the deviation expression is based solely on macro-characteristics (the entries of the elasticity tensor), it may nevertheless be able to provide information on the micro-structure of the material.
Thus, the question arises:

{\it How can the deviation of the elastic properties of natural materials from the ideal Cauchy model be expressed in a unique covariant form?}  

This question brings us back to the irreducible decomposition of tensors. To highlight the issue, we recall an alternative splitting of the elasticity tensor into two independent terms, $M^{ijkl}$ and $N^{ijkl}$, as in (\ref{dev1}). Despite the fact that the skew-symmetric part $N^{ijkl}$ has 6 independent components, its co-partner  $M^{ijkl}$ has a total of 21 separate components. The latter number is lowered to 15 only when the Cauchy relations are precisely met. As a result, the tensor $N^{ijkl}$ is unable to explain the elasticity tensor's deviation from the Cauchy model.

The $S_4$-irreducible decomposition 
\begin{equation}
      C^{ijkl}=S^{ijkl}+A^{ijkl}\,,
\end{equation}
with the sub-tensors given in 
Eqs.(\ref{el-Cauch-part}) and (\ref{el-non-Cauch-part}) fulfills all necessary algebraic requirements:  
    (i) The sub-tensors satisfy the symmetries of the elasticity tensor,
    (ii) Their dimensions are $21=15+6$ as required by the direct sum decomposition of the vector space, and
   (iii) The corresponding subspaces are orthogonal to one another, $S^{ijkl}A_{ijkl}=0$. 

As a result, the tensors $S^{ijkl}$ and $A^{ijkl}$ may be referred to as the Cauchy and non-Cauchy components of the elasticity tensor, respectively. In other words, the deviation of the elasticity tensor from the Cauchy model is expressed by the fourth-order tensor $A^{ijkl}$ or, equivalently, by the second-order tensor $\Delta_{ij}$. 
In \cite{Haus}, a numerical value of the deviation of the elasticity tensor from the Cauchy model was given using some specific entries of the matrix $\Delta_{ij}$. There was also an attempt to attach a physical significance to the sign of this deviation. 
Because the value (and sign) of a specific component of $\Delta_{ij}$ is not retained when a basis is transformed, such a consideration is obviously non-invariant. 
However, it is helpful in the simple (high symmetry) cases of  cubic crystals and isotropic materials.

In order to give an invariant meaning to the deviation from Cauchy's model we apply the Frobenius (Euclidean) norm  of the elasticity tensor 
\begin{equation}
    ||C||=\sqrt{C^{ijkl}C_{ijkl}}\,.
\end{equation}
For the partial tensors $S^{ijkl}$ and $A^{ijkl}$, the norm is expressed similarly,  
\begin{equation}
  ||S||=\sqrt{S^{ijkl}S_{ijkl}}\,, \quad {\rm and} \quad    ||A||=\sqrt{A^{ijkl}A_{ijkl}}\,.
\end{equation}
The scalar $||A||$ presents an invariant expression for the deviation of a material from the ideal Cauchy model. Moreover, we can define a dimensionless {\it Cauchy factor} for an elastic material
\begin{equation}\label{C-factor}
    {\cal F}_{\rm Cauchy} =\frac{||S||}{||C||}=\frac{||S||}{\sqrt{||S||^2+||A||^2}}=\sqrt{\frac{S^{ijkl}S_{ijkl}}{C^{ijkl}C_{ijkl}}}\,.
\end{equation}
Consequently, $0 \leq {\cal F}_{\rm Cauchy} \leq  1.$ A 
pure Cauchy material is determined by ${\cal F}_{\rm Cauchy} = 1.$ 
From the energy consideration, the Cauchy part is essential (non-zero) for every material, thus ${\cal F}_{\rm Cauchy} >0.$ A material with a higher Cauchy factor must have a microscopic structure that is closer to spherical symmetry. The deviation factors for isotropic and cubic materials are shown in \cite{quadr}.  

 Hauss\"uhl \cite{Haus} also proposed to use    the sign of  deviation from the Cauchy model as a characteristic of the microstructure of materials. In particular, the positive deviations  are related to the predominantly ionic bonds, while  the negative deviation is related to the directional bonding, in particular, with strong  covalent bonds. 
 The deviation can be  characterized in invariant form  by a simple scalar parameter $A$ presented in (\ref{A-scalar}). Its sign is well-defined and invariant under the coordinate transformations.  This representation can be seen as a covariant realization of  Hauss\"uhl's proposal. In particular, it is expressed in the division of elastic materials into two classes: $^+A$-materials and  $^-A$-materials. The scalar $A$ is a unique expression for deviation from the Cauchy model when the partial Cauchy relations, $A\ne 0,\, Q_{ij}=0$, hold.

\subsection{Hooke's law decomposed}
Let us recall Hooke's law 
\begin{equation}\label{Hooke3}
    \sigma^{ij}=C^{ijkl}\varepsilon_{kl}.
\end{equation} 
and its diagrammatic presentation, Fig. 1. 
\begin{figure}[ht]
\begin{center}
    \includegraphics[width=0.68\textwidth]{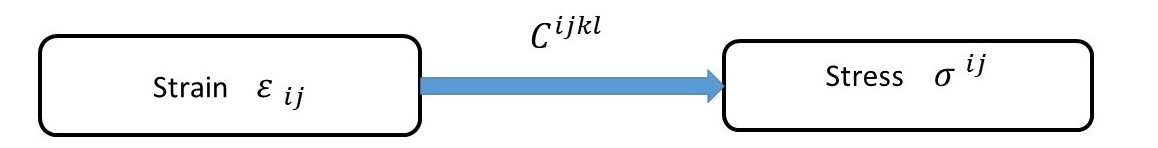}
\caption{ Anisotropic Hooke's Law.}
    \label{fig:Hooke1}
\end{center}
\end{figure}

In this section, we are looking for how the irreducible decomposition of the elasticity tensor shows up in Hooke's law. In this instance, Hooke's law is reduced to a system of two invariant equations,  where the invariant components of the elasticity tensor are represented in distinct ways.

The rotation group allows for an invariant decomposition of the second-order strain tensor $\varepsilon_{ij}$ into two separate parts.
\begin{equation}\label{strain-decomp0}
    \varepsilon_{ij}=\frac 13 \varepsilon g_{ij}+u_{ij},\qquad {\rm where}\quad \varepsilon=g^{ij}\varepsilon_{ij},\quad {\rm and}\quad g^{ij}u_{ij}=0\,. 
\end{equation}
Here, the first term is the {\it mean strain tensor (hydrostatic compression)}. It is proportional to the trace of the strain tensor $\varepsilon$.   
The second traceless part, $u_{ij}$, is referred to as the {\it strain deviator (shear strain)}. Due to  the traceless condition, this term is defined unique. 

Similar to this,  the second-order stress tensor $\sigma^{ij}$ is decomposed irreducibly into two parts---the trace and the traceless tensors:
\begin{equation}\label{stress-decomp}
    \sigma^{ij}=\frac 13 \sigma g^{ij}+s^{ij},\qquad {\rm where}\quad \sigma=g_{ij}\sigma^{ij}, \quad {\rm and}\quad g_{ij}s^{ij}=0\,. 
\end{equation}
The first term here is referred to as the {\it mean (hydrostatic) stress}, 
 while the second term is called the {\it stress deviator (stress shear)}. 
 The decompositions of strain and stress are substituted in Hooke's law (\ref{Hooke3}) to give the following:
 \begin{equation}
     \frac {\sigma}3 g^{ij}+s^{ij}=\frac {\varepsilon}3  C^{ijkl}g_{kl}+C^{ijkl}u_{kl}
 \end{equation}
 To extract the equations for the scalar $\sigma$ and the deviator $s^{ij}$, we first compute the trace on both sides of this equation.
  Thus we have
  \begin{equation}\label{91}
     \sigma =\frac {\varepsilon}3  C^{ijkl}g_{ij}g_{kl}+C^{ijkl}g_{ij}u_{kl}
 \end{equation}
 To calculate the first term of (\ref{91}), we substitute the decomposition (\ref{el-tens-decomp}) and apply the following  traceless identities: 
\begin{eqnarray}
^{(2)}\!S^{ijkl} g_{ij}g_{kl}={}^{(3)}\!S^{ijkl} g_{ij}g_{kl}={}^{(2)}\!A^{ijkl} g_{ij}g_{kl}=0\,.
\end{eqnarray}    
 Thus we obtain
\begin{equation}
\frac {\varepsilon}3  C^{ijkl}g_{ij}g_{kl}=\frac {\varepsilon}3  \left({}^{(1)}S^{ijkl} g_{ij}g_{kl}+{}^{(1)}\!A^{ijkl} g_{ij}g_{kl}\right)=\frac {\varepsilon}3(S+A)\,.
    \end{equation}
For the second term in (\ref{91}), we apply the traceless identities:
\begin{equation}
    {}^{(1)}S^{ijkl}g_{ij}u_{kl}= {}^{(3)}S^{ijkl}g_{ij}u_{kl}=0\qquad{\rm and}\qquad {}^{(1)}A^{ijkl}g_{ij}u_{kl}=0.
\end{equation}
The two remaining expressions are then presented in the form
  \begin{equation}
   {}^{(2)}S^{ijkl}g_{ij}u_{kl}=P^{kl}u_{kl}\qquad{\rm and}\qquad{}^{(2)}A^{ijkl}g_{ij}u_{kl}=-Q^{kl}u_{kl}.
 \end{equation}
As a result, the scalar part of Hooke's law for the mean value of stress is
   \begin{equation}\label{mean-Hooke}
     \sigma = \frac {S+A} 3\varepsilon +(P^{kl}-Q^{kl})u_{kl}.
 \end{equation}
 
 The shear part is extracted from (\ref{91}) as
 \begin{equation}\label{shear-Hook}
     s^{ij}=\frac {\varepsilon}3  C^{ijkl}g_{kl}+C^{ijkl}u_{kl}-\frac {\sigma}3 g^{ij}
 \end{equation}
We apply the relations
 \begin{equation}
     {}^{(1)}S^{ijkl}g_{kl}=\frac S3g^{ij},\qquad  {}^{(2)}S^{ijkl}g_{kl}=P^{ij},\qquad  {}^{(3)}S^{ijkl}g_{kl}=0,
 \end{equation}
 and 
  \begin{equation}
     {}^{(1)}A^{ijkl}g_{kl}=\frac A3g^{ij},\qquad  {}^{(2)}A^{ijkl}g_{kl}=-Q^{ij}.
 \end{equation}
 Consequently, the first term in the right-hand side of (\ref{shear-Hook}) reads
 \begin{equation}\label{first-term}
\frac {\varepsilon}3 C^{ijkl}g_{kl}=
\left(\frac {A+S}9 g^{ij} +\frac{P^{ij}-Q^{ij}}3\right)\varepsilon.
 \end{equation}
 For the second term in the right-hand side of (\ref{shear-Hook}),  we first derive the relations 
 \begin{equation}
     {}^{(1)}S^{ijkl}u_{kl}=\frac {2S}{15}u^{ij},\quad  {}^{(2)}S^{ijkl}u_{kl}=\frac 17 \left(2P^{ik}u^j{}_k+2P^{jk}u^i{}_k+P^{kl}u_{kl}g^{ij}\right),\quad  {}^{(3)}S^{ijkl}u_{kl}=R^{ijkl}u_{kl},
 \end{equation}
 and 
  \begin{equation}
     {}^{(1)}A^{ijkl}u_{kl}=-\frac A6 u^{ij},\qquad  {}^{(2)}A^{ijkl}u_{kl}=Q^{ik}u^j{}_k+Q^{jk}u^i{}_k-Q^{kl}u_{kl}g^{ij}.
 \end{equation}
 Consequently,
 \begin{equation}
C^{ijkl}u_{kl}=\frac {4S-5A}{30}u^{ij} +\left(\frac 27P^{ik}+Q^{ik}\right)u^j{}_k+\left(\frac 27P^{jk}+Q^{jk}\right)u^i{}_k +\left(\frac 17 P^{kl}-Q^{kl}\right)u_{kl}g^{ij}+R^{ijkl}u_{kl}.    
 \end{equation}
As a result, the shear part of the stress tensor is expressed as 
 \begin{eqnarray}\label{shear-Hooke1}
s^{ij}&=&\frac {P^{ij}-Q^{ij}}3\varepsilon+
\frac {4S-5A}{30}u^{ij}+R^{ijkl}u_{kl}+\nonumber\\
&&\frac 27 \left(P^{in}g^{jm} +P^{jn}g^{im}-\frac 23P^{mn}g^{ij}\right)u_{mn}+
\left(Q^{in}g^{jm} +Q^{jn}g^{im}-\frac 23Q^{mn}g^{ij}\right)u_{mn}.
 \end{eqnarray}
 Consequently, the system of two equations (\ref{mean-Hooke}) and (\ref{shear-Hooke1}) is equivalent to the Hooke's law (\ref{Hooke3}). Fig. 3 depicts a graphical representation of this system. The arrows represent the  left-hand side terms of the equations with regard to the invariant irreducible components of the elasticity tensor. 
 \begin{figure}[ht]
\begin{center}
    \includegraphics[width=0.68\textwidth]{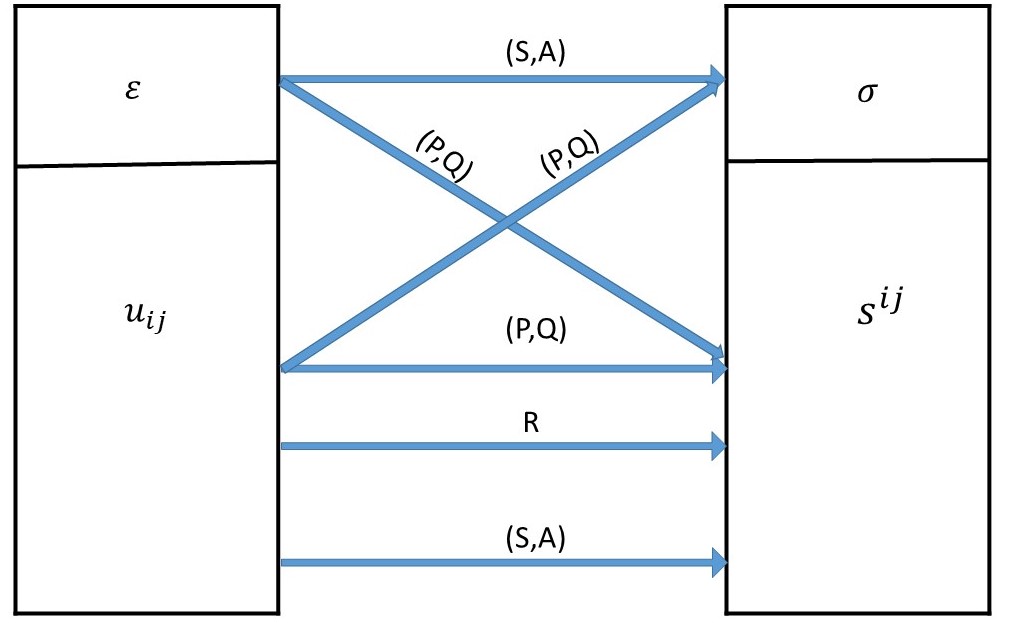}
\caption{ Anisotropic Hooke's Law decomposed. }
    \label{fig:Hooke2}
\end{center}
\end{figure}
 Consider several consequences of these equations
 \begin{itemize}
     \item We notice that the scalar terms $S$ and $A$ appear simultaneously. The second-order tensors $P$ and $Q$ are the same. As a result, Hooke's law does not distinguish between Cauchy and non-Cauchy parts of the elasticity tensor.
     \item The pair of scalars $(S,A)$ gives direct relationships between mean stress and mean strain, as well as shear stress and shear strain. As a result, they might be considered essential components of the elasticity tensor. 
     \item The fourth-order $R$-term does not contribute  to the mean-stress equation.  As a result, it can be thought of as a pure shear part of the elasticity tensor. 
     \item The mixed relations between the mean and shear components of the stress and strain are provided by the second-order tensors $(P,Q)$. The mean-value and the shear effects are entirely separated in the particular situation, \begin{equation}
        P^{ij}=Q^{ij}. 
    \end{equation} 
     \end{itemize}
 Let us observe the simplest special case:      
In the case of isotropic media, when the $P$, $Q$, and $R$-terms vanish, we have the decomposed Hooke's law in the form
     \begin{equation}\label{iso-Hooke}
\sigma = \frac {S+A} 3\varepsilon,\qquad 
s^{ij}=\frac {4S-5A}{30}u^{ij}.
 \end{equation}
 
 As a result, the shear value and mean value variables are separated. The analogs of the one-dimensional Hooke's coefficients for mean value and shear are the coefficients 
 \begin{equation}
     K_{\it mean}=\frac {S+A}3\qquad K_{\it shear}=\frac{4S-5A}{30}. 
 \end{equation}
 
The Cauchy scalar invariant $S$ must be positive in order for the positive stress to result in material compression. We notice the following interesting facts regarding the non-Cauchy scalar invariant $A$:
\begin{itemize}
    \item When applied to materials with positive $A$, the non-Cauchy portion raises the $K_{\it mean}$ coefficient while lowering the $K_{\it shear}$ coefficient. 
     \item Materials with negative $A$ have the opposite effect.
\end{itemize}
We describe how this effect manifests in the energy of the deformed material in the next section. 
\section{Cauchy relations and elasticity energy}
In this section, we discuss the contributions of the irreducible parts of $C^{ijkl}$ to the elasticity energy of deformed material. We are especially interested in how to distinguish between the Cauchy and non-Cauchy contributions. 
\subsection{Three parts of the elasticity energy}
In accordance with Hooke's law, the  quadratic form  of the strain tensor variables represents the elasticity energy,
\begin{equation}\label{energy}
    E=\frac 12\sigma^{ij}\varepsilon_{kl}= \frac 12 C^{ijkl}\varepsilon_{ij}\varepsilon_{kl}\,.
\end{equation}
The major symmetry  of the elasticity tensor $C^{ijkl}=C^{klij}$ follows namely from the latter expression. 
We begin with the conventional decomposition of the strain tensor into two independent parts
\begin{equation}\label{strain-decomp}
    \varepsilon_{ij}=\frac 13 \varepsilon g_{ij}+u_{ij}\, 
\end{equation}
in order to characterize the contributions of the various components of the elasticity tensor into (\ref{energy}). We substitute (\ref{strain-decomp}) into the energy expression (\ref{energy}). As a result, the elastic energy is divided into three components
    \begin{equation}
    E=E_c+E_m+E_s\,.
    \end{equation}
Here,   the  compression part of the energy $E_c$  depends of  the mean strain tensor only, 
\begin{equation}\label{com-energy}
    E_c=\frac 1{18} C^{ijkl}\varepsilon^2g_{ij}g_{kl}\,.
    \end{equation}
 The   shear part of the energy $E_s$ depends of the shear strain only, 
\begin{equation}\label{shear-energy}
E_s=\frac 12 C^{ijkl}u_{ij}u_{kl}\,.
\end{equation}
The mixed part $E_m$ is influenced by both the compression and shear components of the strain, 
\begin{equation}\label{mix-energy}
   E_m=\frac 13 C^{ijkl}\varepsilon g_{ij}u_{kl}\,.
\end{equation}
The decomposition of the elasticity tensor into Cauchy and non-Cauchy parts can be substituted now in these three energy expressions. We use the representation given in  \cite{Itin-MMS}.

\subsection{The compression part of the energy}
By substituting the decomposition (\ref{el-tens-decomp}) into the compression part of the energy (\ref{com-energy}), we have
\begin{equation}
{}^{(c)}E=\frac 1{18} S^{ijkl}\varepsilon^2 g_{ij}g_{kl}+
\frac 1{18} A^{ijkl}\varepsilon^2 g_{ij}g_{kl}\,.
    \end{equation}
The traceless properties yield 
\begin{eqnarray}
^{(2)}\!S^{ijkl} g_{ij}g_{kl}={}^{(3)}\!S^{ijkl} g_{ij}g_{kl}={}^{(2)}\!A^{ijkl} g_{ij}g_{kl}=0\,.
\end{eqnarray}    
Hence, only the scalar invariants of the elasticity tensor contribute to the compression part of the energy. The latter is given as a sum of Cauchy and non-Cauchy parts
\begin{equation}
    E_c={}^{(C)}\!E_c+{}^{(nC)}\!E_c\,.
\end{equation}
The two energy expressions here are determined by the scalars $S$ and $A$, respectively. Indeed, 
 \begin{equation}   
   ^{(C)}\!E_c =\frac {\varepsilon^2S}{18}\,,\qquad {\rm and} \qquad  ^{(nC)}\!E_c=\frac {\varepsilon^2A}{18}\,.
\end{equation}
The total compression part of the energy is expressed now as
\begin{equation}
     E_c=\frac {\varepsilon^2}{18}(S+A)\,.
\end{equation}
Assuming that a material's hydrodynamic compression can be accomplished physically without shear deformation, this energy expression must be positive for any non-zero  $\varepsilon$.   Consequently, we have an inequality 
\begin{equation}
    S+A>0\,.
\end{equation}

\subsection{The mixed part of the energy}
Using the decomposition (\ref{el-tens-decomp}), we can  express the mixed part of the energy (\ref{mix-energy}) as a sum of Cauchy and non-Cauchy parts
\begin{equation}
    E_m={}^{(C)}\!E_m+{}^{(nC)}\!E_m =\frac 13  S^{ijkl}\varepsilon g_{ij}u_{kl}+\frac 13  A^{ijkl}\varepsilon g_{ij}u_{kl}\,.
\end{equation}
Substituting here the decomposition (\ref{S-decomp}) of the Cauchy part into three rotational invariant pieces we derive
\begin{equation}\label{S-energy-decomp}
    ^{(C)}E_m=\frac \varepsilon 3 \left( ^{(1)}\!S^{ijkl}g_{ij}u_{kl}+ ^{(2)}\!S^{ijkl}g_{ij}u_{kl}+ ^{(3)}\!S^{ijkl}g_{ij}u_{kl}\right)
\end{equation}
Since the pure shear $u_{ij}$ is traceless, the first term vanishes.  
The third term includes a completely traceless part $^{(3)}\!S^{ijkl}g_{ij}=0$, thus it likewise vanishes. As a result, all that is left is the contribution of the second component of the elasticity tensor. Explicitly, 
\begin{equation}
    ^{(C)}E_m=\frac {\varepsilon }3\,   ^{(2)}\!S^{ijkl}g_{ij}u_{kl}=
    \frac {\varepsilon}3\, P^{kl}u_{kl}\,.
\end{equation}

The decomposition of the non-Cauchy part of the elasticity tensor  into the rotation invariant terms yields 
\begin{equation}
    ^{(nC)}E_m =\frac {\varepsilon}3 \left( ^{(1)}\!A^{ijkl}g_{ij}u_{kl}+{}^{(2)}\!A^{ijkl}g_{ij}u_{kl}\right)
\end{equation}
The first term in this equation is zero due to the tracelessness of the shear strain $u_{kl}$. As for the second term, we have 
\begin{equation}
^{(nC)}E_m =-\frac{\varepsilon}3 Q^{kl}u_{kl}.
\end{equation}
The total mixed energy is expressed as
\begin{equation}
    E_m =\frac {\varepsilon}3(P^{kl}-Q^{kl})u_{kl}\,.
\end{equation}
  Consequently, only two second-order deviators—out of the elasticity tensor's five irreducible components—\\contribute to the mixed energy. The mixed portion of the energy disappears if these two terms are not present. Both isotropic materials and cubic crystals exhibit this behavior. These two cases were discussed in the literature; see \cite{Rychlewski1}, \cite{Rychlewski2}. We find that the mixed energy can disappear even in less symmetric materials  when the particular relation  $P^{kl}=Q^{kl}$ between the components is valid.

\subsection{The shear part of the energy}
The quadratic form of shear strain  energy given in (\ref{shear-energy})  is decomposed into the Cauchy and non-Cauchy parts
\begin{equation}
    E_s= ^{(C)}\!E_s+ ^{(nC)}\!E_s=\frac 12 S^{ijkl}u_{ij}u_{kl}+\frac 12 A^{ijkl}u_{ij}u_{kl}\,.
\end{equation}
All three irreducible sub-tensors, $^{(1)}\,S_{ijkl}, 
{}^{(2)}\,S_{ijkl}, $ and $^{(3)}\,S_{ijkl}$,  contribute to the Cauchy portion of the shear energy.
 Explicitly, we have the first part 
\begin{equation}
   {}^{(1)}\!S^{ijkl}u_{ij}u_{kl}=
  \frac {2S}{15}\,u_{ij}u^{ij}\,,
\end{equation}
the second part 
\begin{equation}
 {}^{(2)}\!S^{ijkl}u_{ij}u_{kl}=
 \frac 47\,P^{ik}u^j{}_ku_{ij}\,,
\end{equation}
and the third part
\begin{equation}
   {}^{(3)}\!S^{ijkl}u_{ij}u_{kl}=
   R^{ijkl}u_{ij}u_{kl}\,.
\end{equation}
Consequently, the Cauchy part of the shear energy reads
\begin{equation}
    ^{(C)}\!E_s=\frac {S}{15}\,u_{ij}u^{ij}+
    \frac 27\,P^{ik}u_{ij}u^j{}_k+\frac 12\, R^{ijkl}u_{ij}u_{kl}\,.
\end{equation}
The non-Cauchy part of the shear energy also includes contributions of all its  irreducible subtensors $^{(1)}A_{ijkl}, $ and $^{(2)}A_{ijkl}$.
Respectively, 
\begin{equation}
    {}^{(1)}\!A^{ijkl}u_{ij}u_{kl}=
    -\frac {A}{6}\,u_{ij}u^{ij}\,,
\end{equation}
and
\begin{equation}
    {}^{(2)}\!A^{ijkl}u_{ij}u_{kl}=2Q^{ik}u^j{}_ku_{ij}\,.
\end{equation}
As a result, the non-Cauchy part of the shear elasticity energy is expressed as 
\begin{equation}
    ^{(nC)}\!E_s=-\frac {A}{12}\,u_{ij}u^{ij}+
    Q^{ij}u_{jk}u^k{}_i\,.
\end{equation}
Fig.3 illustrates the elasticity energy decomposition graphically. Take into account that each of the three energy portions is contributed to by both the Cauchy and non-Cauchy parts of the elasticity tensor.
\begin{figure}[ht]
\includegraphics[width=1.0\textwidth]{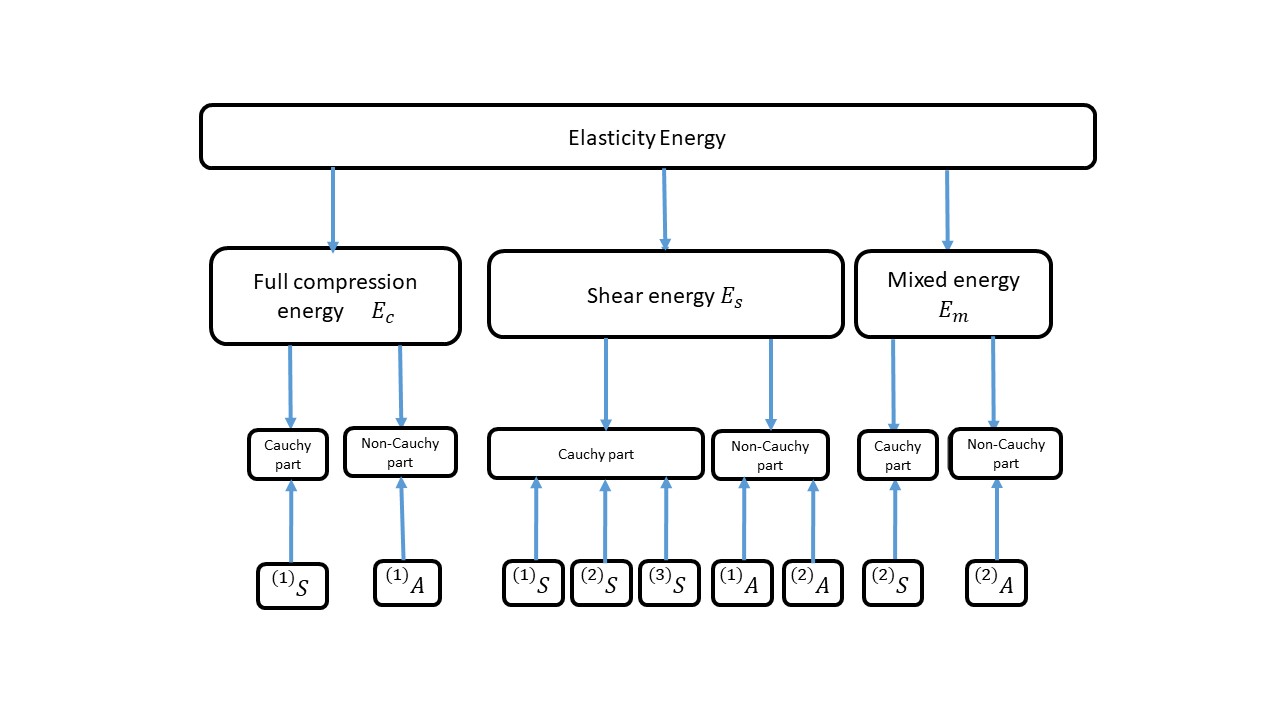}
\caption{ $E_c,E_s$, and $E_m$ are three independent elements that make up the elasticity energy. Each of these pieces is broken down into Cauchy and non-Cauchy components. The contributions of irreducible subtensors of $C^{ijkl}$   are identified.}
    \label{fig:mesh1}
\end{figure}
\subsection{Cauchy and non-Cauchy energy contributions}
To explicitly characterize the distinction between the Cauchy and non-Cauchy contributions to the energy expression, we consider the most basic case of an isotropic medium.
In this example, the mixed portion ${}^{(m)}E$ is not present.
Cauchy and non-Cauchy portions of the compression energy ${}^{(c)}E$ are separated as 
\begin{equation}\label{contrib1}
    {}^{(c)}\!E=\frac {S}{18}\varepsilon^2+\frac {A}{18}\varepsilon^2\,,
\end{equation}
respectively. 
The shear energy also decomposed into Cauchy and non-Cauchy parts  
\begin{equation}\label{contrib2}
    {}^{(s)}\!E=\frac S{15}(u_{ij})^2-\frac {A}{12}\,(u_{ij})^2\,.
\end{equation}
We can infer the following facts from these formulas: 
\begin{itemize}
    \item Since pure shear and compression can both be achieved independently, we  demand that the compression and shear components of the energy be positive independently. 
    As a result of Eqs.(\ref{contrib1},\ref{contrib2}), the scalar invariant $S$ must be strictly positive. This suggests that the Cauchy component is crucial, making pure non-Cauchy materials with $S=0, A\ne 0$ impossible.
    \item With a positive parameter $A$, the non-Cauchy portion contributes in the opposite direction to the energy expressions, increasing hydrostatic compression energy while decreasing pure shear energy. The contributions are inverted when the parameter $A$ is negative. With the use of this property, we are able to physically discriminate between the Cauchy and non-Cauchy components of the elasticity tensor. It is important to note that this interesting property is only disclosed by the irreducible scalar invariants $S$ and $A$. When the reducible decomposition with Lam\'e's parameters $\lambda,\mu$ is applied, it is not visible.
    \item Positiveness of the full energy requires 
    \begin{equation}
        S+A>0,\qquad 4S-5A>0\,.
    \end{equation}
    Note that these inequalities must hold  only for a positive parameter $S$. Consequently, the non-Cauchy scalar $A$ is bounded as 
    \begin{equation}\label{SA-bound}
        0.8S>A>-S
    \end{equation}
    In term of Poisson's ratio, this inequality identical to the classical thermodynamics bound \cite{Landau}
    \begin{equation}
        0.5>\nu>-1.
    \end{equation}
\end{itemize}

\section{Acoustic waves and Cauchy relations}
In this section, we look at how the Cauchy and non-Cauchy components of the elasticity tensor affect acoustic wave propagation properties. 
\subsection{Acoustic wave equation} 
The propagation of acoustic waves in anisotropic media is given by a system of three second-order partial differential equations:
\begin{equation}\label{wave-eq}
{  \rho g^{il}\ddot{u_l}-C^{ijkl}\,{\partial_j\partial_k u_l}=0\,.}
\end{equation}
Here, the displacement covector $u_l=u_l(t, x^i)$ is assumed to be a
smooth function of the time coordinate $t$ and of the spatial position $x^i$ of a point. The mass density $\rho$, the elasticity tensor $C^{ijkl}$, and the metric tensor $g^{il}$ are treated as constant parameters.  Moreover, we use Cartesian coordinates, so the Euclidean metric is presented by the unit matrix $g^{ij}=\text{diag}(1,1,1)$.

We consider a  {\it plane wave solution} of Eq.  (\ref{wave-eq}), with the
notation as in 
\cite{Nayfeh}:
\begin{equation}\label{wave-an}
u_l=U_le^{i\left(\zeta n_jx^j-\omega t\right)}\,.
\end{equation}
Here,   $U_i$ is  the  amplitude covector, $\zeta$ is the wave-number, $n_j$ is the unit propagation 
covector, $\omega$ is the angular  frequency, and $i^2=-1$. In the 
plane-wave approximation, all these parameters are assumed to be constant. Then, by substituting  (\ref{wave-an})  into (\ref{wave-eq}), a system of three homogeneous linear algebraic equations is obtained,
\begin{equation}\label{wave-an1}
\left(\rho\, \omega^2g^{il}-C^{ijkl}\zeta^2n_jn_k\right)U_l=0\,.
\end{equation}
It has a non-trivial solution if and only if the characteristic matrix is singular. As a result, 
\begin{equation}\label{char}
\det\left(\rho \omega^2g^{il}-C^{ijkl}\zeta^2n_jn_k\right)=0\,.
\end{equation}
\subsection{Christoffel tensor}
With the definitions of the {\it Christoffel tensor} \begin{equation}\label{Christoffel}
\Gamma^{il}:=\frac 1\rho\, C^{ijkl}n_jn_k\,
\end{equation}
and of the phase velocity $v:=\omega/\zeta$, 
Eq. (\ref{wave-an1}) becomes an ordinary eigenvalue system 
\begin{equation}\label{char1}
\left( v^2g^{il}-\Gamma^{il}\right)U_l=0\,,
\end{equation}
{whereas} the characteristic equation (\ref{char}) reads
\begin{equation}\label{char2}
\det\left( v^2g^{il}-\Gamma^{il}\right)=0\,.
\end{equation}
Due to the minor and major symmetries of the elasticity tensor (\ref{el-sym1},\ref{el-sym2}), the Christoffel tensor turns out to be symmetric
\begin{equation}\label{Christoffel1}
{\Gamma^{ij}=\Gamma^{ji}\,.}
\end{equation}
Consequently, the eigenvalues $v_1^2,v_2^2,v_3^2$ are real. Furthermore, these eigenvalues must be positive, implying that the symmetric matrix $\Gamma^{ij}$ must be positive-definite.

 Due to the dependence of the matrix $\Gamma^{ij}$ on $n_i$, each positive real solution $v_i$ corresponds to an acoustic wave propagating in the direction of the wave covector $n_i$. Thus, in general, for a given propagation covector ${\mathbf n}=(n_1,n_2,n_3)$, three different acoustic waves are possible.  The relations below are worth observing: 
\begin{equation}\label{sum-vel2}
    v_1^2+v_2^2+v_3^2={\rm tr}(\Gamma^{ij}),
\end{equation}
and 
\begin{equation}
    v_1v_2v_3=\sqrt{{\rm det}(\Gamma^{ij})}.
\end{equation}
The cases with
zero solutions (null modes) and negative solutions (standing waves)
are classified as nonphysical since they do not satisfy the standard causality requirements. As a result, non-degenerate wave propagation is only possible in the directions ${\mathbf n}$ that meet the relations 
\begin{equation}\label{G-ineq}
    {\rm tr}\left(\Gamma^{ij}({\mathbf n})\right)>0,\qquad  {\rm det}\left(\Gamma^{ij}({\mathbf n})\right)>0. 
\end{equation}

\subsection{$S_4$-decomposition of  Christoffel's tensor}
Christoffel's tensor can be separated into Cauchy and non-Cauchy parts since it is linear in the elasticity tensor.
Substituting the irreducible $S\!A$-decomposition into
(\ref{Christoffel}), we define 
\begin{equation}\label{Chr-dec1}
\Gamma^{il}={\cal S}^{il}+{\cal A}^{il}\,,
\end{equation}
where
\begin{equation}\label{Chr-dec2}
  {\cal S}^{il}:=\frac 1\rho\, S^{ijkl}n_jn_k\,,\qquad {\rm and }\qquad
  \A^{il}:=\frac 1\rho\,A^{ijkl}n_jn_k  \,.
\end{equation}
These two tensors are symmetric due to the fundamental symmetries of the elasticity tensor, 
\begin{equation}
 {\cal S}^{il}={\cal S}^{li},\qquad{\rm and}\qquad  {\cal A}^{il}={\cal A}^{li}.  
\end{equation}
 They correspond to the Cauchy and non-Cauchy parts of the elasticity tensor. Therefore, we shall refer to ${\cal S}^{ij}$ as the {\it Cauchy Christoffel tensor} and to ${\cal A}^{ij}$ as the {\it non-Cauchy Christoffel tensor.} 
  {Substituting} (\ref{el-Cauch-part}) and
(\ref{el-non-Cauch-part}) into (\ref{Chr-dec2}), we find the explicit expressions
\begin{equation}\label{Chr-dec3}
{\cal S}^{il}=\frac 1{3\rho}(C^{ijkl}+C^{iklj}+ C^{iljk})n_jn_k=\frac 1{3\rho}(2C^{ijkl}+ C^{iljk})n_jn_k\,
\end{equation}
and
\begin{equation}\label{Chr-dec4}
{\cal A}^{il}=\frac1{3\rho}\left(2C^{ijkl}-C^{iklj}-C^{ilkj}\right)n_jn_k=\frac1{3\rho}
\left(C^{ijkl}-C^{ilkj}\right)n_jn_k\,.
\end{equation}
We can learn something valuable from the latter expression. 
 The non-Cauchy part of the Christoffel tensor  satisfies the relation 
\begin{equation}\label{A-rel}
    {\cal A}^{il}n_l=0\,.
\end{equation}
Indeed, if we contract the covector $n_l$ with ${\cal A}^{il}$, we  obtain using the symmetry relation (\ref{el-non-Cauch-ident}) 
 \begin{equation}
     {\cal A}^{il}n_l=A^{ijkl}n_jn_kn_l=A^{i(jkl)}n_jn_kn_l=0\,.
 \end{equation}
Eq. (\ref{A-rel}) can be viewed as a linear relation between the {rows} of the matrix $\A^{il}$. 
Consequently, this matrix is singular. Thus we derive that the determinant of the non-Cauchy  Christoffel tensor is equal to zero,
\begin{equation}\label{detA}
\det\,{\cal A}^{ij}=0\,.
\end{equation}
This relation demonstrates that a medium with only a non-Cauchy part ${\cal A}^{ij}$  is impossible since the second inequality in (\ref{G-ineq}) yields zero propagation velocity. We come to the conclusion that wave propagation in any elastic material requires the Cauchy portion of the Christoffel tensor ${\cal S}^{ij}$ to be non-zero.

\subsection{$SO(3)$-decomposition of Christoffel's tensor}
The invariance of the Christoffel tensor under the action of the rotation group $SO(3)$ allows for its successive decomposition. 
Let us first compute the Christoffel tensor contributions of the three irreducible parts that compose the fully symmetric tensor $S^{ijkl}$. 
For the scalar part (\ref{S1}), we have
\begin{equation}
   ^{(1)}\!{\cal S}^{il}=\frac 1\rho\, ^{(1)}\!S^{ijkl}n_jn_k=\frac{S}{15\rho}(g^{il}+2n^in^l). 
\end{equation}
The trace of this tensor is positive
\begin{equation}
   ^{(1)}\!{\cal S}^{il}g_{il}=\frac{S}{3\rho}
\end{equation}
Applying (\ref{S2}), the second part $^{(2)}\!S^{ijkl}$ is given as 
\begin{equation}
   ^{(2)}\!{\cal S}^{il}=\frac 1\rho\, ^{(2)}\!S^{ijkl}n_jn_k=\frac 1{7\rho}\left(P^{il}+g^{il}P^{jk}n_jn_k+2P^{il}n_jn^l+2P^{lj}n_jn^i\right)
\end{equation}
with the trace 
\begin{equation}
   ^{(2)}\!{\cal S}^{il}g_{il}=\frac 1\rho\,P^{ij}n_in_j\,.
\end{equation}
The third part contributes into the Christoffel tensor as 
\begin{equation}
   ^{(3)}\!{\cal S}^{il}=\frac 1\rho\, ^{(3)}\!S^{ijkl}n_jn_k=\frac 1\rho R^{ijkl}n_jn_k\,.
\end{equation}
Due to the traceless properties of $R^{ijkl}$, the trace of this tensor is zero
\begin{equation}
   ^{(3)}\!{\cal S}^{il}g_{il} =R^{ijkl}n_jn_kg_{il}=0\,.
\end{equation}
As a result, the Cauchy part of the Christoffel tensor is presented as 
\begin{equation}\label{S-cristof}
    {\cal{S}}^{il}=\frac{S}{15\rho}(g^{il}+2n^in^l)+\frac 1{7\rho}\left(2P^{ij}n_jn^l+2P^{lj}n_jn^i+P^{il}+P^{jk}n_jn_kg^{il}\right)+\frac 1\rho R^{ijkl}n_jn_k\,.
\end{equation}
The trace of this expression is expressed as 
\begin{equation}
    {\rm tr}({\cal{S}}^{ij})=\frac{S}{3\rho}+\frac 1\rho\,P^{ij}n_in_j.
\end{equation}
It is worth noting that the tensor $R^{ijkl}$ (of 9 independent components of the elasticity tensor) does not contribute to the latter expression.

Let us compute now the non-Cauchy contributions into Christoffel's tensor. The scalar part contribution  is given by applying (\ref{A1-tens}) 
\begin{equation}
  ^{(1)}\!{\cal A}^{il}= \frac 1\rho\, ^{(1)}\!A^{ijkl}n_jn_k =
\frac{A}{12\rho} \left(n^in^l-g^{il}\right)
\end{equation}
with the trace 
\begin{equation}
  ^{(1)}\!{\cal A}^{il}g_{il}= 
-\frac{A}{6\rho}. 
\end{equation}
This contribution is positive for the  $A^-$-materials and negative for the $A^+$-materials. 

The contribution of the second non-Cauchy part can be calculated using (\ref{A2-g}). We have  
\begin{equation}
  ^{(2)}\!{\cal A}^{il}= \frac 1\rho\, ^{(2)}\!A^{ijkl}n_jn_k  =\frac 1{2\rho}\left(Q^{il}+g^{il}Q^{jk}n_jn_k-Q^{ik}n^ln_k-Q^{lk}n^in_k\right)
\end{equation}
with the trace 
\begin{equation}
  ^{(2)}\!{\cal A}^{il}g_{il}=\frac 1{2\rho}Q^{ij}n_in_j\,.
\end{equation}
Consequently, the non-Cauchy contribution to Christoffel's tensor takes the form
\begin{equation}\label{A-cristof}
  {\cal A}^{il}=\frac{A}{12\rho} \left(n^in^l-g^{il}\right)+\frac 1{2\rho}\left(Q^{il}+g^{il}Q^{jk}n_jn_k-Q^{ik}n^ln_k-Q^{lk}n^in_k\right)  
\end{equation}
with the trace 
\begin{equation}\label{nC-Cris}
    {\rm tr}({\cal A}^{il})=-\frac{A}{6\rho}+\frac 1{2\rho}Q^{ij}n_in_j\,.
\end{equation}
Finaly, Christoffel's tensor is represented in terms of the irreducible parts of the elasticity tensor as 
\begin{eqnarray}\label{Gamma}
\Gamma^{il}&=&\frac{S}{15\rho}(g^{il}+2n^in^l)+\frac 1{7\rho}\left(2P^{ij}n_jn^l+2P^{lj}n_jn^i+P^{il}+P^{jk}n_jn_kg^{il}\right)+\frac 1\rho R^{ijkl}n_jn_k+\nonumber\\
&&\frac{A}{12\rho} \left(n^in^l-g^{il}\right)+\frac 1{2\rho}\left(Q^{il}+g^{il}Q^{jk}n_jn_k-Q^{ik}n^ln_k-Q^{lk}n^in_k\right).
\end{eqnarray}
The trace of this expression is given by 
\begin{equation}
    {\rm tr}(\Gamma^{ij})=g_{ij}\Gamma^{ij}= 
   \frac{2S-A}{6\rho}+\frac 1{2\rho}\,\left(2P^{ij}+Q^{ij}\right)n_in_j.
\end{equation}

\subsection{Acoustic velocities}
Let us consider certain consequences of the  Christoffel's tensor represented in terms of the irreducible parts of the elasticity tensor.

\subsubsection{ Sum of squared velocities} 
The second coefficient of the characteristic equation that is represented by the trace of the matrix ${\rm tr}(\Gamma^{il})={\rm tr}({\cal S}^{il}+{\cal A}^{il})$ is equal to the sum of the eigenvalues. 
Thus, using (\ref{sum-vel2}) we have for a given propagation direction ${\mathbf{n}}$ the sum of three squared velocities
\begin{equation}\label{sum-vel2-exp}
    \sum_{a=1}^3v_a^2({\mathbf{n}})=
   \frac{2S-A}{6\rho}+\frac 1{2\rho}\,(2P^{ij}+Q^{ij})n_in_j
\end{equation}
The preceding expression takes a much simpler  form in the special case 
\begin{equation}\label{2P-Q}
    2P^{ij}+Q^{ij}=0.
\end{equation}
The sum of the squared velocities of three separate waves propagating in the direction ${\mathbf{n}}$ is then {\it independent} of the propagation vector  ${\mathbf{n}}$ 
\begin{equation}\label{sum-vel2-exp0}
    \sum_{a=1}^3v_a^2({\mathbf{n}})=
   \frac{2S-A}{6\rho}\,.
\end{equation}
Using the bound (\ref{SA-bound}), we obtain that in this special case
\begin{equation}
    0.2\,\frac S\rho<\sum_{a=1}^3v_a^2({\mathbf{n}})<
    0.5\,\frac S\rho\,.
\end{equation}
Thus the velocities of three acoustic waves are bounded by the Cauchy scalar invariant $S$ only. Once more, we can observe that this invariant has to be strictly positive $S>0$. 

Notice that  only five linear constraints between the 21 independent components of $C^{ijkl}$ are required in Eq. (\ref{2P-Q}). 
As a result, a large class of elasticity tensors with 16 arbitrary components can exhibit this essential property (\ref{sum-vel2-exp0}). 
In particular, this case is  realized for cubic crystals and isotropic materials.

\subsubsection{ Critical directions for the sum of squared velocities}
The sum of squared velocities in anisotropic media generally depends on the propagation vector ${\mathbf{n}}$. 
Let us look for the directions where this sum achieves its minimum and maximum values. We will  refer to these directions as {\it critical directions}. 
Let us write the function given in the right-hand side of (\ref{sum-vel2-exp}) as
\begin{equation}\label{K(s)}
    K({\mathbf{n}})=\frac{2S-A}{6\rho}+\frac 1{2\rho}\,L^{ij}n_in_j,
\end{equation}
where $L^{ij}=2P^{ij}+Q^{ij}$. 
The critical values of this expression on the sphere  $|{\mathbf{n}}|=1$ are given by the characterictic equation
\begin{equation}
   L^{ij}n_j=\lambda n^i,
\end{equation}
where $\lambda$ is Lagrangian's multiplier. As a result, the critical directions ${\mathbf{n}}$ are the eigenvector of the matrix $L^{ij}$, and the parameter $\lambda$ is the eigenvalue of this matrix.

\subsubsection{Sum of squared velocities for three orthogonal  direction} 
Let us assume that the three orthonormal vectors ${\mathbf{n}},{\mathbf{m}},{\mathbf{k}}$ represent three separate propagation directions.
The components of these vectors are connected by a well-known relation 
\begin{equation}
    n_in_j+m_im_j+k_ik_j=g_{ij}.
\end{equation}
Using three copies of equation (\ref{sum-vel2-exp}) for three orthogonal directions we obtain the well-known invariant expression, see e.g. \cite{Alshits}, that in our notations reads 
\begin{equation}\label{sum-vel2-exp1}
    \sum_{a=1}^3\big(v_a^2({\mathbf{n}})+v_a^2({\mathbf{m}})+v_a^2({\mathbf{k}})\big)=
   \frac{2S-A}{2\rho}\,.
\end{equation}
Thus, for every elastic material, the sum of the squared norms of the three propagation velocities in every three orthogonal directions depends only on the scalar invariants of the elasticity tensor. In particular, it is independent of the propagation vectors.  This amount ranges from $0.6S/\rho$ to $1.5S/\rho$.
\subsubsection{Acoustic velocities in isotropic media and cubic system}
It is instructive to apply the formulas for the Christoffel tensor in the isotropic material. Three tensors, $P^{ij}$, $Q^{ij}$, and $R^{ijkl}$ are disappearing in this instance.
The Christoffel tensor then takes the form  
\begin{equation}
\Gamma^{ij}=\frac{4S-5A}{60\rho}g^{ij}+\frac{8S+5A}{60\rho}n^in^j\,.
\end{equation}
As a result, characteristic equation (\ref{wave-an1}) becomes
\begin{equation}
{\det}\left(\left(v^2-\frac{4S-5A}{60\rho}\right)g^{ij}-\frac{8S+5A}{60\rho}n^in^j\right)=0.
\end{equation}
Instead of directly solving this cubic equation with respect to velocity $v^2$, we can deduce the velocities using simple algebraic reasons. 
We observe that this equation can be rewritten as the standard characteristic equation, ${\det}\left(\lambda g_{ij}-n_in_j\right)=0$.  Since the matrix $n_in_j$ has a rank of one this equation has two zero eigenvalues. There are therefore two equal velocities,
\begin{equation}
   v_1^2=v_2^2=\frac{4S-5A}{60\rho}.
\end{equation}
The third velocity is derived from (\ref{sum-vel2-exp1}) as
\begin{equation}
    v_3^2=\frac{2S-A}{6\rho}-2v_1^2=\frac S{5\rho}.
\end{equation}
In term of the Lam\'e moduli $\lambda$ and $\mu$, we come back to the well-known textbook's expressions
\begin{equation}
   v_1^2=v_2^2=\frac \mu\rho,\qquad v_3^2=\frac{\lambda+2\mu}\rho.
\end{equation}
The velocity $v_3$ in the isotropic case therefore solely depends on the Cauchy part of the elasticity tensor. The possibility to generalize this fact to a more general anisotropic case will be discussed in the following section.

The tensors $P^{ij}$ and $Q^{ij}$  disappear in cubic system crystals, while $R^{ijkl}$ only has one independent component. Now, the Christoffel tensor is dependent on the propagation vector as
\begin{equation}
    \Gamma^{ij}=\frac{4S-5A}{60\rho}g^{ij}+\frac{8S+5A}{60\rho}n^in^j+R^{ijkl}n_kn_l.
\end{equation}
The trace of this expression gives exactly the same sum of squared velocities as in the isotropic case. 
\subsubsection{Critical directions for some crystal systems}
In trigonal, tetragonal, and transverse isotropy systems, the tensors $P^{ij}$ and $Q^{ij}$ can be represented simultaneously in the form proportional to the matrix ${\rm diag}(1,1,-2)$; see \cite{Itin-MMS} for explicit formulae. 
Consequently, the matrix $L^{ij}$ appearing in (\ref{K(s)}) is given by
\begin{equation}
    L^{ij}=(2P+Q){\rm diag}(1,1,-2).
\end{equation}
The explicit expressions for the scalars $P,Q$ can be find in \cite{Itin-MMS}. Then the eigenvalues are $\lambda_1=\lambda_2=1$ while $\lambda_3=-2$. The standard axes serve as the eigenvectors. 
Then in two orthogonal directions ${\mathbf{n}_1}=(1,0,0)$ and ${\mathbf{n}_2}=(0,1,0)$ (corresponding to the eigenvalue $\lambda=1$), the sum of squared velocities takes the value 
\begin{equation}
    \sum_{a=1}^3v_a^2({\mathbf{n}_1})= \sum_{a=1}^3v_a^2({\mathbf{n}_2})=
   \frac{2S-A}{6\rho}+\frac 1{2\rho}(2P+Q)
\end{equation}
while in the third direction ${\mathbf{n}_3}=(0,0,1)$ (corresponding to the the eigenvalue $\lambda=-2$) this sum is given by
\begin{equation}
    \sum_{a=1}^3v_a^2({\mathbf{n}_3})=
   \frac{2S-A}{6\rho}-\frac 1{\rho}(2P+Q).
\end{equation}
We can derive from these equations a bound for the scalars $P$ and $Q$
\begin{equation}
\frac{2S-A}{6}>2P+Q>-\frac{2S-A}{3}.
\end{equation}
Using the energy bound (\ref{SA-bound}) we have  
\begin{equation}
  0.5S>2P+Q>-S.
\end{equation}
Although these inequalities are only derived in a special case, they highlight an essential fact: the tensor parts of the elasticity tensor are not fully arbitrary.
  They are constrained by the scalar part parameters. 

\subsection{Polarization of acoustic waves}
The propagation of acoustic waves in a homogeneous anisotropic elastic media is represented by equation (\ref{char1}).
The three eigenvalues of this eigenvector problem indicate the velocities of three distinct acoustic waves.
In general, three distinct real positive eigenvalues correspond to three distinct eigenvectors called {\it acoustic polarizations}: $^{(1)}\mathbf{U}$, $^{(2)}\mathbf{U}$, and $^{(3)}\mathbf{U}$.

For isotropic materials, the three polarization vectors  $^{(i)}\mathbf{U}$ are strongly correlated with the propagation vector $\mathbf{n}$. 
One {\it longitudinal {or compression} wave} exhibits polarization along the propagation vector i.e.,
\begin{equation}\label{long}
 \mathbf{U}\times\mathbf{n}=0\,. 
\end{equation}
Two additional {\it {transverse (or shear)  waves}} have polarizations perpendicular to the propagation direction, i.e.,
\begin{equation}\label{shear}
	  \mathbf{U}\cdot\mathbf{n}=0\,.
\end{equation}
These three {\it pure polarizations} can be found in any direction of the vector $\mathbf{n}$. 

In anisotropic material, three pure modes  only arise in specific directions of $\mathbf{n}$. It is critical to identify these pure modes and the wave covector directions that correspond to them. Let us look at how the irreducible decomposition of the elasticity tensor can be utilized in this context.

\subsubsection{Pure longitudinal wave} 
Let in a given direction $\mathbf{n}$ the polarization be pure longitudinal, i.e., the polarization vector  is proportional to the wave covector, $^{(1)}\mathbf{U}\sim \mathbf{n}$. 
Two additional waves must have polarization vectors $^{(2)}\mathbf{U}$ and $^{(3)}\mathbf{U}$ that are normal to the wave covector $\mathbf{n}$, i.e. they are pure transverse.
 For the pure longitudinal wave, we can replace $^{(1)}\mathbf{U}$ with $\mathbf{n}$.  Thus, Eq. (\ref{char1}) becomes  
\begin{equation}\label{char1-x}
\Gamma^{ij}n_j=v_L^2g^{ij}n_j\,.
\end{equation}
By substituting the decomposition of the Christoffel tensor into Cauchy and non-Cauchy parts, we obtain
\begin{equation}\label{char1-xy}
{\cal S}^{ij}n_j+{\cal A}^{ij}n_j=v_L^2g^{ij}n_j\,.
\end{equation}
Because of Eq. (\ref{A-rel}), the second term on the left-hand side is omitted, leaving us with 
\begin{equation}\label{char1-xyz}
{\cal S}^{ij}n_j=v_L^2g^{ij}n_j\,.
\end{equation}
The velocity $v_{\rm{L}}$ of the pure longitudinal wave in the direction of $\mathbf{n}$ is then determined solely by the Cauchy part of the elasticity tensor: 
\begin{equation}\label{long-vel}
v^2_L={{\cal S}^{ij}n_jn_i}\,.
\end{equation}
By substituting this expression into Eq. (\ref{char1-xyz}), we obtain 
\begin{equation}\label{pure-cond}
    {\cal S}^{ij}n_j=\left({{\cal S}^{rs}n_rn_s}\right) n^i. 
\end{equation}
Being a vector equation for a single vector variable, it allows for the computation of pure longitudinal wave directions ${\mathbf n}$ in a material with a specified elasticity tensor. This direction is then independent of the non-Cauchy part of the elasticity tensor. Furthermore, because two extra pure shear waves are always normal to a pure longitudinal wave, we can regard Eq. (\ref{pure-cond}) as a {\it condition for the existence of three pure waves} in a certain direction of $\mathbf{n}$.

In isotropic media, the condition (\ref{pure-cond}) is trivially satisfied for any arbitrary $\mathbf{n}$. Moreover, according to this equation, even some anisotropic materials can have three pure polarizations in any direction.  For this, the elasticity tensor must have an isotropic Cauchy part, whereas the non-Cauchy part might be arbitrary.

Let us explicitly express the longitudinal velocity in terms of elasticity tensor components. By substituting (\ref{S-cristof}) in (\ref{long-vel}), we obtain
\begin{equation}\label{long-vel1}
v^2_{\rm{L}}={\frac 1{\rho}\left( {\frac{S}{15}+\frac 67 P^{ij}n_in_j +R^{ijkl}n_in_jn_kn_l}\right)}\,.
\end{equation}
 Consequently, for the two pure shear velocities we have 
\begin{equation}
    v_{S1}^2+v_{S2}^2=\frac 1{\rho}\left(\frac{8S-5A}{30}+\frac 1{14}\,(2P^{ij}+7Q^{ij})n_in_j- R^{ijkl}n_in_jn_kn_l\right).
\end{equation}
Two shear velocities are equal one to another in the case of an acoustic axes \cite{Norris},
 $v_{S1}=v_{S2}$. In this case, the shear velocity is given by
\begin{equation}
    v_S^2=\frac 1{2\rho}\left(\frac{8S-5A}{30}+\frac 1{14}\,(2P^{ij}+7Q^{ij})n_in_j- R^{ijkl}n_in_jn_kn_l\right). 
\end{equation}
Hence, the shear velocities are affected by the Cauchy and non-Cauchy parts of the elasticity tensor.

\subsection{Pure transverse wave}
Consider a more general case whereby a pure transverse wave ${\mathbf U}$ propagates in a certain direction ${\mathbf n}$.
 Then, ${\mathbf U}\cdot {\mathbf n}=0$. Two more waves are normal to ${\mathbf U}$ but can have any angle with the vector ${\mathbf n}$.
When we multiply both sides of the equation \begin{equation}\label{shear-eq}
    \Gamma^{ij}U_j=v^2_Sg^{ij}U_j
\end{equation}
by the vector ${\mathbf n}$, we obtain
\begin{equation}
    {\cal S}^{ij}U_jn_i=0.
\end{equation}
where the relation $\Gamma^{ij}n_i={\cal S}^{ij}n_i$ is applied.
Thus, we are dealing with a system of two scalar equations
\begin{equation}\label{pure-cond2}
{\cal S}^{ij}n_iU_j=0\qquad {\rm and} \qquad  g^{ij}n_iU_j=0
\end{equation}
for two independent components of the shear polarization vector ${\mathbf U}$. 
We suppose that they are not pure longitudinal waves, i.e.,  $\Gamma^{ij}n_i\ne Cg^{ij}n_i$.  
The shear polarization ${\mathbf U}$ is then uniquely determined by our system of equations.  To resolve the system (\ref{pure-cond2}) explicitly, we note that this system can be interpreted as the fact that the vector ${\mathbf U}$ is orthogonal to two given vectors, ${\cal S}^{ij}n_i$ and $g^{ij}n_i$. As a result, the pure shear polarization vector can be expressed as a vector product 
\begin{equation}\label{shear-vector}
    U_i=U_0\varepsilon_{ijk}n^j{\cal S}^{kl}n_l\,,
\end{equation}
In addition to the explicit expression, we discover that the vector ${\mathbf U}$ is independent of the non-Cauchy part of the elasticity tensor when only one (shear) polarization appears.

With the polarization vector in hand, we can compute the {\it shear velocity}.
 Multiplying both sides of   Eq. (\ref{shear-eq}) by $U_j$, we obtain
\begin{equation}\label{shear-velocity}
    v_S^2=\frac{\Gamma^{ij}U_iU_j}{|U|^2}. 
\end{equation}
This quantity is determined by the full Christoffel's tensor.

Substituting (\ref{shear-vector}) into (\ref{shear-eq}), we derive 
\begin{equation}\label{shear-condition}
    \varepsilon_{imk}\left(\Gamma^{ij}-v_S^2g^{ij}\right){\cal S}^{kl}n^m n_l=0\,.
\end{equation}
There is only one independent variable in this system, ${\mathbf n}$. As a result, it can be regarded as a {\it condition  for the presence of the pure shear wave} in the direction ${\mathbf n}$.

Let us provide an explicit example. Consider the propagation vector to be directed in the $z$-axis, ${\mathbf n}=(0,0,1)$ and the metric tensor $g_{ij}={\rm diag}(1,1,1)$.
 Then Eq. (\ref{shear-vector}) yields 
\begin{equation}
U_i=U_0\left(S_{13},S_{23},0\right).
\end{equation}
The shear velocity (\ref{shear-velocity}) is then given as 
\begin{equation}
    v_S^2=\frac {\Gamma_{11}S_{13}^2+2\Gamma_{12}S_{13}S_{23}+\Gamma_{22}S_{23}^2}{S_{13}^2+S_{23}^2}.
\end{equation}
The shear conditions (\ref{shear-condition}) are presented as a system
\begin{eqnarray}
\Gamma_{11}S_{23}-\Gamma_{12}S_{13}&=&v^2_S S_{23}\nonumber\\
\Gamma_{22}S_{13}-\Gamma_{12}S_{23}&=&v^2_S S_{13}\nonumber\\
A_{13}S_{23}-A_{23}S_{13}&=&0\,.
\end{eqnarray}

\section{Conclusion}
The Cauchy relations of linear elasticity are examined from a pure algebraic perspective. 
\begin{itemize}
    \item We demonstrate that the Cauchy relations are related to the permutation group symmetry of the elasticity tensor. The linear space of the elasticity tensor is then uniquely split into Cauchy and non-Cauchy subspaces.
    \item The Cauchy relations are expressed as the condition for nullification of the 6-dimensional non-Cauchy subspace of the elasticity tensor space.
    \item We argue that almost arbitrary invariant linear relation between elasticity tensor components is  equivalent to the system Cauchy relations.
    \item We propose identifying Cauchy and non-Cauchy parts of the elasticity tensor since natural materials do not exactly meet Cauchy relations.
    \item The Cauchy and non-Cauchy subspaces are separated into smaller invariant subspaces by successive $SO(3,\mathbb R)$-group decomposition.
    \item The $SO(3,\mathbb R)$-decomposition of the non-Cauchy subspace allows for the definition of partial Cauchy relations of five independent conditions. Unlike the full Cauchy conditions, these partial conditions hold for isotropic materials as well as cubic system crystals.
    \item We define two types of materials: the $A^+$-materials and the $A^-$-materials.
\end{itemize}
 We analyze the physical consequences of the Cauchy relations in the following aspects:
\begin{itemize}
    \item The deviation of natural materials from the ideal Cauchy model is described in an invariant form. We also provide some examples of the $A^+$ and $A^-$ materials. 
    \item The decomposition of Hooke's law enables the identification of the contribution of various components of the elasticity tensor. We find that the $A^+$ and $A^-$ materials behave differently.
    \item We analyze the partitioning of elasticity energy into compression, shear, and mixed portions. The contributions of the invariant sub-tensors of the elasticity tensor to distinct energy expressions are detected.
    \item We derive some constraints on the invariant elasticity parameters from the positiveness of the energy.
    \item We present a unique decomposition of the acoustic Christoffel tensor into Cauchy and non-Cauchy portions for acoustic wave propagation.
    \item The contributions of the irreducible elements of the elasticity tensor to the velocities and propagation vector of pure polarized acoustic waves are identified. The pure longitudinal velocity is demonstrated to be only reliant on the Cauchy part of the elasticity tensor. When there is only one shear wave, its polarization vector is independent of the non-Cauchy part.
\end{itemize}

\section*{Acknowledgements} 
F.-W. Hehl (Cologne), R.L. Fosdick (Minnesota), V. I. Alshits (Moscow), M. Vianello (Milan), A.H. Norris (Rutgers), M.B. Rubin (Haifa), and R. Segev (Beer Sheva) deserve my heartfelt thanks. I would want to thank all of the elasticity and acoustics experts with whom I have had extremely useful discussions (both online and in person) about various aspects of the Cauchy relations over the years. I appreciate the reviewers' high level of expertise and their extremely helpful recommendations.  

\section*{Conflict of interest statement} The author has no conflicts of interest to declare.


\begin{thebibliography}{99}
 \bibitem{Alshits} Alshits, V. I., \& Lothe, J. (2004). Some basic properties of bulk elastic waves in anisotropic media. {\it Wave Motion} {\bf 40}(4), 297-313.

\bibitem{Backus} Backus, G.  (1970). A geometrical picture of anisotropic elastic tensors. {\it Rev. Geophys. Space Phys.} {\bf 8}, 633--671.

\bibitem{Baerheim} Baerheim, R. (1993). Harmonic decomposition of the anisotropic elasticity tensor. {\it Quarterly J.\ Mech.\ Appl.\
    Math.} {\bf 46}, 391--418.
        
 \bibitem{Bona}   B\'{o}na, A., Bucataru, I., \& Slawinski, M. A. (2004). Material symmetries of elasticity tensors. {\it Quarterly J.\ Mech.\ Appl.\
    Math.} {\bf 57}, 583-598.

   \bibitem{Cowin1989} Cowin, S.~C. (1989). Properties of the anisotropic elasticity tensor. {\it Quarterly J.\ Mech.\ Appl.\ Math.} {\bf 42}, 249--266. Corrigenda  ibid. (1993) {\bf 46}, 541--542.

\bibitem{Cowin1992} Cowin, S.~C.\ \& Mehrabadi, M.~M. (1992). The structure of the linear anisotropic elastic symmetries. {\it J.\ Mech.\ Phys.\ Solids} {\bf 40}, 1459--1471.

\bibitem{Cowin1995a}  Cowin, S.~C.\ \& Mehrabadi, M.~M. (1995).  Anisotropic symmetries of linear elasticity. {\it Appl. Mech. Rev.} {\bf 48}(5), 247--285.


\bibitem{Desmorat} Desmorat, R., Auffray, N., Desmorat, B., Olive, M., \& Kolev, B. (2021). Minimal functional bases for elasticity tensor symmetry classes. {\it Journal of Elasticity} {\bf 147}(1-2), 201-228.

\bibitem{Elcoro2010} Elcoro, L., \& Etxebarria, J. (2010). Common misconceptions about the dynamical theory of crystal lattices: Cauchy relations, lattice potentials and infinite crystals. {\it European journal of physics} {\bf 32}(1), 25.  

\bibitem{Epstein1946} Epstein, P. S. (1946). On the elastic properties of lattices.  {\it Physical Review} {\bf 70}(11-12), 915.

\bibitem{Forte1} Forte, S., \& Vianello, M. (1998). Functional bases for transversely isotropic and transversely hemitropic invariants of elasticity tensors. {\it Quarterly J. Mech. Appl. Math.} {\bf 51}(4), 543-552.

\bibitem{Forte2} Forte, S., \& Vianello, M. (1996). Symmetry classes for elasticity tensors. {\it Journal of Elasticity} {\bf 43}(2), 81-108.

\bibitem{Hamermesh} Hamermesh, M. (1969) {\it Group Theory and its Application to Physical Problems. } New York: Dover Publications.


\bibitem{Fosdick}  Fosdick, R.L. (May 2002). Private communications.


\bibitem{Haus} Hauss\"uhl, S. (2007). {\it Physical Properties of Crystals: An Introduction.} Weinheim, Germany: Wiley-VCH.

\bibitem
{Hehl-Obukhov(2003)}
Hehl, F.~W.\ \& Obukhov, Yu.~N. (2003). {\it  Foundations of
    Classical Electrodynamics.}
  Boston, MA: Birkh\"auser.

\bibitem{Cauchy1}
 Hehl, F.~W.\ \& Itin, Y. (2002).  The Cauchy relations in linear elasticity theory.  {\it Journal of Elasticity} {\bf 66}, 185--192.
  
\bibitem{Cauchy2}
  Itin, Y.,\ \&  Hehl, F.~W. (2013). The constitutive tensor of linear elasticity: Its decompositions, Cauchy relations, null Lagrangians, and wave propagation. {\it J. Math. Phys.} {\bf 54}, 042903.

\bibitem{quadr} Itin, Y. (2016). Quadratic invariants of the elasticity tensor. {\it Journal of Elasticity} {\bf 125}(1), 39-62.

\bibitem{Itin-MMS} Itin, Y. (2020). Irreducible matrix resolution for symmetry classes of elasticity tensors. {\it Mathematics and Mechanics of Solids} {\bf 25}(10), 1873-1895.

\bibitem{IR} Itin, Y., \& Reches, S. (2022). Decomposition of third-order constitutive tensors. {\it Mathematics and Mechanics of Solids} {\bf 27}(2), 222-249.



\bibitem{Kaxiras} Kaxiras, E. (2003) {\it  Atomic and Electronic Structure of Solids}. Cambridge: Cambridge University Press.

 \bibitem{Landau} Landau, L. D., \& Lifshitz, E. M. (1986) {\it Theory of Elasticity}, 3rd. ed. Oxford, UK:  Pergamon Press.


 \bibitem{Lancia} Lancia, M. R., Caffarelli, G. V., \& Podio-Guidugli, P. (1995). Null Lagrangians in linear elasticity. {\it Mathematical Models and Methods in Applied Sciences} {\bf 5}(04), 415-427.

 
  \bibitem{Love} Love, A.~E.~H. (1944). {\it  A Treatise on the Mathematical Theory of Elasticity}, 4th ed. New York: Dover Publications. 

\bibitem{MacDonald1972} MacDonald, R. A. (1972). Cauchy relations for second-and third-order elastic constants. {\it Physical Review B  } {\bf 5}(10), 4139.

\bibitem{Marsden} Marsden, J.~E.\ \& Hughes, T.~J.~R. (1983). {\it
    Mathematical Foundations of Elasticity.} Englewood Cliffs, NJ: Prentice-Hall.

\bibitem{Mochizuki} Mochizuki, E. (1988). Spherical harmonic decomposition of an elastic tensor. {\it Geophysical Journal International} {\bf 93}(3), 521-526.

\bibitem{Mott} Mott, P. H., \& Roland, C. M. (2013). Limits to Poisson's ratio in isotropic materials—general result for arbitrary deformation. {\it Physica Scripta} {\bf 87}(5), 055404.


\bibitem{Nayfeh} Nayfeh, A.~H. (1985) {\it Wave Propagation in Layered Anisotropic Media: with Applications to Composites.} Amsterdam: North-Holland.
  
  
\bibitem{Norris}  Norris, A. N. (2004). Acoustic axes in elasticity. {\it Wave Motion} {\bf 40}(4), 315-328.

\bibitem{Nye} Nye, J. F. (1985). {\it Physical Properties of Crystals: their Representation by Tensors and Matrices.} Oxford: Oxford University Press.

\bibitem{Olive} Olive, M., Kolev, B., Desmorat, R., \& Desmorat, B. (2022). Characterization of the symmetry class of an elasticity tensor using polynomial covariants. {\it Mathematics and Mechanics of Solids} {\bf 27}(1), 144-190.


\bibitem{Perrin} Perrin, B. (1979). Cauchy relations revisited. {\it Phys. Stat. Sol. B } {\bf 91}, K115–K120.



\bibitem{Podio} Podio-Guidugli, P. (2000). A Primer in Elasticity. {\it Journal of Elasticity} {\bf 58}, 1–104. Reprinted (2013). The Netherlands: Kluwer Academic Publisher.

\bibitem{Podio1} Podio-Guidugli, P. (2020). On null-Lagrangian energy and plate paradoxes. In: Altenbach, H., Chinchaladze, N., Kienzler, R., Müller, W. (eds) {\it Analysis of Shells, Plates, and Beams: Advanced Structured Materials,} vol 134, 367-372. Switzerland: Springer.




\bibitem{Rychlewski1} Rychlewski, J. (2000). A qualitative approach to Hooke's tensors. Part I. {\it Archives of Mechanics } {\bf 52}(4-5), 737-759.

\bibitem{Rychlewski2} Rychlewski, J. (2001). A qualitative approach to Hooke's tensors. Part II. {\it Archives of Mechanics } {\bf 53}(1), 45-63.

\bibitem{Rubin} Rubin, M. B., \& Ehret, A. E. (2018). Invariants for Rari-and Multi-Constant Theories with Generalization to Anisotropy in Biological Tissues. {\it Journal of Elasticity} {\bf 133}(1), 119-127.

\bibitem{Stakgold} Stakgold, I. (1950). The Cauchy relations in a molecular theory of elasticity. {\it Quarterly of Applied Mathematics} {\bf 8}(2), 169-186.

\bibitem{Sirotin} Sirotin, Y. (1975). Decomposition of material tensors into irreducible parts. {\it Sov. Phys. Crystallogr.} {\bf 19}, 565-568.

\bibitem{Sirotin-book} Sirotin, Y. I., \& Shaskol’skaya, M. P. (1979). {\it Principles of Crystal Physics.} Moscow: Nauka.

\bibitem{Zener1947} Zener, C. (1947). A defense of the Cauchy relations. {\it Physical Review} {\bf  71}(5), 323.

\bibitem{Weyl} Weyl, H. (2016). {\it The Classical Groups.} Princeton, NJ: Princeton University Press.













\end{thebibliography}
\end{document}